\documentclass[epj]{svjour}
\usepackage[british]{babel}
\usepackage{epsfig}
\usepackage{appendix}
\usepackage{subfig}
\usepackage[T1]{fontenc}

\usepackage[switch,columnwise]{lineno}

\usepackage[space]{cite}
\usepackage{amsmath}
\usepackage{multirow}

\newcommand\ba{\begin{eqnarray}}
\newcommand\ea{\end{eqnarray}}
\newcommand{\be}{\begin{equation}}
\newcommand{\ee}{\end{equation}}
\newcommand{\M} {{\cal M}} 

\newcommand\bi{\begin{itemize}}
\newcommand\ei{\end{itemize}}
\newcommand\bc{\begin{center}}
\newcommand\ec{\end{center}}

\usepackage{xspace}
\newcommand\nn{\nonumber}
\newcommand\eq[1]{\begin{align} #1 \end{align}}
\newcommand\ga[1]{\begin{gather} #1 \end{gather}}
\newcommand{\br}[1]{\left( #1 \right)}
\newcommand{\brs}[1]{\left[ #1 \right]}
\newcommand{\brm}[1]{\left| #1 \right|}
\newcommand{\Li}[2]{\mbox{Li}_{#1}\!\br{#2}}
\newcommand{\GeV}{\mbox{GeV}}
\newcommand{\Sp}{\mbox{Tr}}
\newcommand{\fb}{\mbox{fb}}
\newcommand{\pb}{\mbox{pb}}
\newcommand{\dd}[1]{{\hat #1}}   % ë¢•‡‚™† · gamma-¨†‚‡®Ê•© Ñ®‡†™†.

\newcommand{\vv}[1]{{\bf #1}}
\renewcommand{\Re}{\mbox{Re}}
\usepackage{tikz}
\usetikzlibrary{decorations.markings}
\usetikzlibrary{decorations.pathmorphing}
\usetikzlibrary{patterns}

\tikzset{
    plain/.style={line width=1},
    photon/.style = {line width=0.8,decorate, decoration={snake,amplitude=1pt,segment length=5pt}},
    fermion/.style={line width=1,postaction={decorate}, decoration={markings,mark=at position #1 with {\arrow{stealth}}}},
    fermion/.default=.55,
    neutrino/.style={line width=1,dotted},
    plain_compound/.style={line width=0.7,double},
    charged_compound/.style={line width=0.7,double,postaction={decorate}, decoration={markings,mark=at position #1 with {\arrow{stealth}}}},
    charged_compound/.default=.55,
    charged_scalar/.style={dashed,line width=0.7,postaction={decorate}, decoration={markings,mark=at position #1 with {\arrow{stealth}}}},
    charged_scalar/.default=.55,
    Z/.style={dashed,line width=0.7},
    W/.style={dashed,line width=0.7,postaction={decorate}, decoration={markings,mark=at position #1 with {\arrow{stealth}}}},
    W/.default=.55,
    electron/.style={postaction={decorate}, decoration={markings,mark=at position #1 with {\arrow{stealth}}}},
    electron/.default=.55,
    gluon/.style={decorate, decoration={coil,amplitude=2pt, segment length=2.5pt}},
    show curve controls/.style={
        decoration={
            show path construction,
            curveto code={
                \draw [blue, dashed]
                    (\tikzinputsegmentfirst) -- (\tikzinputsegmentsupporta)
                    node [at end, cross out, draw, solid, red, inner sep=3pt]{};
                \draw [blue, dashed]
                    (\tikzinputsegmentsupportb) -- (\tikzinputsegmentlast)
                    node [at start, cross out, draw, solid, red, inner sep=3pt]{};
            }
        }, decorate
    },
    vector/.style={->,>={Triangle[scale width=#1]}},
    vector/.default=0.5
}
\tikzset{
  pics/carc/.style args={#1:#2:#3}{
    code={
      \draw[pic actions] (#1:#3) arc(#1:#2:#3);
    }
  }
}

\begin{document}

\title{Radiative corrections in proton--antiproton annihilation to electron-positron and their application to the PANDA experiment}
\subtitle{Radiative corrections to $\bar p p \to e^+e^-$ }
\author{
Yu.M.~Bystritskiy\inst{1}
\and 
V.A.~Zykunov\inst{1,2} 
\and 
A.~Dbeyssi\inst{3}
\and 
M.~Zambrana\inst{3}
\and 
F.~Maas\inst{3} 
\and 
E.~Tomasi-Gustafsson\inst{5}
}
\institute{Joint Institute for Nuclear Research, 141980 Dubna, Russia
\and
Francisk Skorina Gomel State University,  246019 Gomel, Belarus
\and
Helmholtz Institute Mainz, Staudingerweg 18, D-55128 Mainz, Germany
\and
CEA, IRFU, SPhN, Saclay, F-91191 Gif-sur-Yvette, France
}

\date{\today}
%\date{Received: date / Revised version: date}
\abstract{
Radiative corrections to the annihilation of proton--antiproton into electron--positron are revisited,
including virtual and real (soft and hard) photon emission. This issue is relevant for the time-like
form factors measurements planned at the PANDA experiment at the FAIR facility, in next future.
The relevant formulas are given. A stand-alone Monte-Carlo integrator is developed on the basis
of the calculated radiative cross section and its application to the PANDA experiment is illustrated.
\PACS{
      {PACS-key}{discribing text of that key}   \and
      {PACS-key}{discribing text of that key}
     } % end of PACS codes
} %end of abstract
\maketitle

%----------------------------------------------------
\section{Introduction}
\label{Introduction}
%----------------------------------------------------

The elementary annihilation process $\bar{p} + p \to e^+ + e^-$ and the time-reverse reaction 
$e^+ + e^- \to \bar{p} + p$ contain direct information on the proton electromagnetic form factors (FFs).
FFs are fundamental quantities that conveniently parametrize the electric and magnetic currents in the proton.
The kinematical region accessed by annihilation reactions is the time-like (TL) region of transferred momenta 
$q^2$, where $q^2$ is positive. Assuming that the reactions occur through the exchange of a virtual photon of
four momentum squared $q^2$, the annihilation cross section is parametrized in terms of two complex amplitudes,
that are functions of $q^2$ only. At the leading order in the fine electromagnetic constant,
$\alpha = e^2/4\pi \approx 1/137$, (Born approximation), the cross section contains the moduli squared of FFs,
and shows a linear dependence in $\cos^2 \theta$, where $\theta$ is the angle of the produced electron
in the center of mass system (c.m.s). The precise measurement of the angular distribution of one of the final particles
allows to access directly the electric and magnetic FFs \cite{Zichichi:1962ni}.

However, the charged particles involved in the reaction irradiate and the emission of real or virtual photons requires
higher order corrections to the measured cross section in order to recover the Born cross
section (Fig.~\ref{fig.Born}) and extract TL FFs.

Extensive literature is dedicated to radiative corrections to electron proton scattering (for a recent
review see \cite{Pacetti:2015iqa}), few works were previously dedicated to this physics issue in the context
of the physics program of the PANDA \cite{Peters:2017kop} experiment at FAIR \cite{Spiller:2018mta}.
In space-like (SL) region updated radiative corrections calculations were made necessary by
the program of FF measurements at the Jefferson Laboratory, at large transferred momentum. Following
the Akhiezer--Rekalo polarization method \cite{Akhiezer:1968ek,Akhiezer:1974em}, the GEp collaboration measured
precisely the ratio of the electric $G_E$ to magnetic $G_M$ FFs in a series of experiments \cite{Puckett:2017flj}
and found a large deviation from unity, contrary to what previously suggested by unpolarized cross section
measurements, using the Rosenbluth method \cite{Rosenbluth:1950yq}.

Among the possible explanations of this
discrepancy: normalization issues in the data \cite{Pacetti:2016tqi}, correlations in the parameters of
the Rosenbluth fit \cite{TomasiGustafsson:2006pa}, radiative corrections at higher orders \cite{Bystritskiy:2006ju}
or more precise first order corrections \cite{Gramolin:2016hjt}, the community largely focussed on a
possible enhancement of the two photon exchange mechanism \cite{Afanasev:2017gsk}. Model dependent
calculations showed that a large effect increasing with $Q^2 = -q^2$ could be indeed found
\cite{Blunden:2017nby,Guttmann:2010au}. An exact calculation is not feasible, because one should
know the $Q^2$ dependence of the FFs for all intermediate proton excited states. Model independent considerations,
however, predict that the two photon contribution introduces charge odd terms, that are sources of non-linearities in
the Rosenbluth plots, in SL region, and of odd powers of $\cos\theta$ terms in TL region. As for today,
no experimental evidence of enhancement of the two photon exchange contribution has been found (for a recent discussion, see \cite{Bytev:2019rdc}. The importance of
this issue is related to the fact that, in presence of multi-photon exchange, the formalism that relates
the observables to FFs does not hold any more. Instead of two real FFs, function of $Q^2$, the scattering
process would be described by three amplitudes in general complex functions of two kinematical variables
\cite{Rekalo:2003km,Rekalo:2004wa}. It would be still possible to extract the real FFs, functions of $Q^2$
but at the price of difficult measurements of polarization observables, including double and triple
polarizations, of the order of $\alpha$. Attention should be given to this problem, because even a few
percent relative contribution of the two photon mechanism, would bring when neglected large inconsistencies
in the FFs extraction. In the TL region, the two photon exchange contribution could be easier to be measured.
Having a precise angular distribution, the sum (difference) of the differential cross section at complementary
angles cancels (enhances) the charge odd contributions. The sum of cross sections for $\theta$ and $\pi-\theta$
would still be an even function of $\cos\theta$, and of $G_{E,M}^2\br{Q^2}$ \cite{Gakh:2005wa,Gakh:2005hh}.

In the TL domain, precise radiative corrections have been implemented for experiment at electron-positron colliders
at LEP and more recently for BESIII \cite{Czyz:2004ua, Czyz:2014sha}. The necessary high precision is
obtained with Monte-Carlo based on the lepton Structure Function method \cite{Kuraev:1985hb} that allows to take
into account higher orders in the leading logarithmic approximation.

The feasibility studies for the measurement of the $\bar{p}p \to e^+ e^-$ process have been performed
\cite{Sudol:2009vc}, and more recently within the PANDARoot framework \cite{Singh:2016dtf}, at the different
energies accessible by the PANDA experiment. The expected precision on the measurements of $\brm{G_E}$,
$\brm{G_M}$, and on their ratio $R$ were evaluated, assuming an integrated luminosity of $2~\fb^{-1}$ per
beam momentum setting. The results of the simulations show that the proton form factor ratio can be
measured with a total relative uncertainty between $3.3\%$ at $s = 5.08~\GeV^2$ and $57\%$ at $s = 13.9~\GeV^2$,
including both statistical and systematical uncertainties. The relative uncertainty on $\brm{G_E}$ ($\brm{G_M}$)
is between $2.2\%$ ($3.5\%$) and $48\%$ ($9.7\%$). The measurement of total cross section, that gives an
effective form factor, can be in principle extended to higher energy values ($\sim 40~\GeV^2$) depending on the
real experimental efficiency.

The PANDARoot analysis program includes the package PHOTOS \cite{Golonka:2006tw} for the calculation of radiative
corrections, which are not specific to the annihilation reaction $\bar{p} + p \to e^+ + e^-$. In particular the odd
effects that may arise from initial (ISR) and final (FSR) state radiation interference (INT) are not taken into
account. These effects may induce large errors, if neglected, for the extraction of FFs, as they destroy the symmetry
of the angular distribution. Monte-Carlo simulations indicate that the precision of the FFs measurement at PANDA will
be of the order of few percent in the near threshold region, becoming larger at larger energy
\cite{Sudol:2009vc,Singh:2016dtf}. Therefore it is necessary to quantify high order effect although first order
calculations seem sufficient at this stage to be implemented in the data analysis. Radiative corrections (RC) due to
the emission of real and virtual photons do affect the measurement of the experimentally observable quantities, in
particular the differential cross section. The individual determination of the electric and magnetic proton FFs
requires the precise knowledge of the angular distribution of the final lepton, in shape and in absolute value.

The motivation of this paper is to derive and collect the relevant formulas for soft and hard photon emission in
$\bar{p} + p \to e^+ + e^-$ in a convenient form to be implemented in a stand-alone Monte-Carlo integrator and to compare the
results of the simulations with previous calculations.

The paper is organized as follows:
Section~\ref{sec.Born} gives the notations of the four momenta and the definitions
and formulas for the Born cross section. Section~\ref{sec.RC} contains the evaluation of different radiative
contributions to the process. These contributions are presented in the corresponding subsections: 
subsections~\ref{sec.VirtualRC} and \ref{sec.SoftRC} give the expressions for virtual radiative corrections and for
soft real photon emission contribution to the Born diagram and compares to the existing calculations,
while subsection~\ref{sec.HardRC} is devoted to the hard photon emission overview.
Section~\ref{sec.Numerical} gives the results in form of tables and figures, proving the stability of the numbers with
respect to the relevant parameters.
And in Section~\ref{sec.Conclusion} one can find our conclusion on the result obtained.

%----------------------------------------------------
\section{Born cross section}
\label{sec.Born}
%----------------------------------------------------

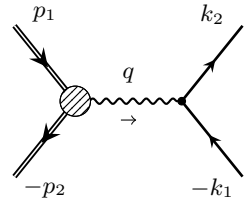
\begin{figure}
\begin{center}
    \begin{tikzpicture}
        % Left proton.
        \draw[charged_compound] (-1.5,1) -- (-0.7,0.0);
        \draw[charged_compound] (-0.7,0.0) -- (-1.5,-1);
        % Right lepton.
        \draw[fermion] (0.7,0.0) -- (1.5,1);
        \draw[fermion] (1.5,-1) -- (0.7,0.0);
        % Intermediate particles.
        \draw[photon] (-0.7,0.0) -- (0.7,0.0);
        % Interaction points.
        \draw[fill=white]                           (-0.7,0.0) circle (0.2);
        \draw[fill=black, pattern=north east lines] (-0.7,0.0) circle (0.2);
        \draw[fill=black] (0.7,0.0) circle (0.05);
        % Momenta.
        \node at (-1.1,1.15) {$p_1$};
        \node at (-1.1,-1.15) {$-p_2$};
        \node at (1.1,1.15) {$k_2$};
        \node at (1.1,-1.15) {$-k_1$};
        \node at (0,0.35) {$q$};
        \draw[->] (-0.1,-0.25) -- (0.1,-0.25);
    \end{tikzpicture}
	\caption{\label{fig.Born} Feynman diagram for the process (\ref{eq.Process}) at tree (Born) level. }
\end{center}
\end{figure}

The basic process is
\eq{
	p(p_1)+\bar{p}(p_2) \to e^+(k_1)+e^-(k_2),
	\label{eq.Process}
}
where the particle four-momenta are indicated in parentheses. It is illustrated in Fig.~\ref{fig.Born}.
The masses of particles are denoted in the following way:
$M$ is mass of proton ($p_1^2=p_2^2=M^2$) and $m$ is mass of electron ($k_1^2=k_2^2=m^2$).
The velocityies of the proton, electron and any fermion $f$ with mass $m_f$ are: 
\ga{
	\beta_p = \sqrt{1-\frac{4M^2}{s}},
	\ 
	\beta_e = \sqrt{1-\frac{4m^2}{s}},
	\ 
	\beta_f = \sqrt{1-\frac{4m_f^2}{s}}.
}
In the kinematical conditions of PANDA the electron mass $m$ can be neglected (i.e.,  $k_1^2 = k_2^2 = 0$ is set 
everywhere possible), then $\beta_e \to 1$, but the proton mass $M$ can not be neglected, therefore, for simplicity, 
the index $p$ in the definition of $\beta_p$ is neglected and $\beta_p$ is replaced with the notation $\beta \equiv \beta_p$ in all
formulas below.

The relativistic kinematics of process (\ref{eq.Process}) is described in terms of Mandelstam invariants:
\eq{
	s &= \br{p_1 + p_2}^2 = 2M \br{M+E}, \label{eq.s}
	\\
	t &= \br{p_2 - k_2}^2 = -\frac{s}{4}\br{1+\beta^2-2\beta \cos\theta}, \label{eq.t}
	\\
	u &= \br{p_1 - k_2}^2 = -\frac{s}{4}\br{1+\beta^2+2\beta \cos\theta}, \label{eq.u}
}
where $\theta$ is the scattering angle between the directions of the antiproton beam and of the final
electron $\vv{k_2}$ in the c.m.s of initial particles,  $E=\sqrt{M^2 + P^2}$ ($P=\brm{\vv{p_2}}$) is the energy(momentum) of the antiproton in the laboratory (Lab) frame (for the proton target:  $\vec p_1=0$). The antiproton beam momentum range in PANDA is $P = 1.5\div15$~\GeV. 
The following relation holds: 
\eq{
	s+t+u=2m^2+2M^2 \approx 2 M^2,  \label{eq.stu}
}
based on the 4-momentum conservation.
In the Born approximation and in  c.m.s. of the initial $p\bar{p}$-pair 
the differential cross section has the form \cite{Zichichi:1962ni}:
\ba
	\sigma_B \equiv \frac{d\sigma_B}{d\Omega_-} &=& 
	\frac{\alpha^2}{4 s\beta}
	\Bigl[ \brm{G_M(s)}^2 \br{1+\cos^2\theta} + \\\nn
        && 
       \br{1-\beta^2} \brm{G_E(s)}^2 \sin^2\theta \Bigr] ,
	\label{eq.BornWithFF}
\ea
where the electric $G_E$ and the magnetic $G_M$ Sachs FFs of the proton \cite{Sachs:1962zzc}
are related to the Dirac and Pauli form factors $F_{1,2}$ by:
\ba 
	G_E(q^2) &= F_1(q^2) + \tau F_2(q^2), \nn\\
	G_M(q^2) &= F_1(q^2) + F_2(q^2),
      \label{eq.SachsFF}
\ea
where $\tau = q^2/4M^2$.
The four momentum transfer at the proton vertex at the Born level is
\eq{
	q = p_1 + p_2,
	\qquad
	q^2 = s.
	\label{eq.q}
}
At zero momentum transfer the electric and magnetic form factors
are normalized respectively to the charge and to the magnetic moment $\mu_p=2.79$ of the proton expressed in
nuclear magnetons,
i.e., $G_E(0) = 1$, $G_M(0) = \mu_p$ and, correspondingly,
the Dirac and Pauli form factors are $F_1(0)=1$, $F_2(0)=\mu_p-1$.
In the assumption of a point-like proton one has:
\eq{
	G_E(q^2) = G_M(q^2) = 1, \ \mbox{or}\  F_1(q^2)=1, \ F_2(q^2)=0,
	\label{eq.PointLikeProton}
}
and the Born cross section then reads as:
\eq{
	\sigma_B^0 \equiv \frac{d\sigma_B^0}{d\Omega_-} = 
	\frac{\alpha^2}{4 s\beta}
	\br{ 2 - \beta^2 \sin^2\theta }.
	\label{eq.PointLikeBorn}
}
Different models and parametrizations exist for the nucleon form factors. We will use either a modified dipole parametrization:
\eq{
	G_E(q^2) &= \frac{M_0^4}{(q^2+M_0^2)^2}, \\
	G_M(q^2) &= \mu_p \, G_E(q^2),
}
where $M_0^2=0.71~\GeV^2$ or the vector dominance model extended to the TL region from \cite{Iachello:2004aq}.

The Born cross section depends on FFs, which parametrization have in general been fitted on the existing data. It
turns out that the results of radiative corrections depend very weakly on the FF model.

%----------------------------------------------------
\section{Radiative corrections}
\label{sec.RC}
%----------------------------------------------------

The radiative corrections to the process (\ref{eq.Process}) at the next to leading order in $\alpha$ include virtual
photon emission such as vertex corrections, vacuum loops and the exchange of two photons, as well as real photon
emission, from initial and final states, and their interferences. The total amplitude $\M_{tot}$
is the incoherent sum of the amplitudes for the two different final states: $e^+ + e^-$ including the Born
diagram $\M_B$ and the virtual corrections $\M_V$, and $e^+ + e^- + \gamma$ describing the real photon emission
$\M_\gamma$ in the initial (ISR) and final (FSR) states:
\ba
	\sigma
	&\sim&
	\brm{\M_{tot}}^2
	=
	\brm{\M_B + \M_V}^2
	+
	\brm{\M_\gamma}^2\nn\\
	&=&
	\brm{\M_B}^2
	+
	2 \,\Re\br{ \M_B^+ \, \M_V }
	+
	\brm{\M_V}^2
	+
	\brm{\M_\gamma}^2,
\label{eq.eqm}
\ea
where the amplitudes $\M_V$ and $\M_\gamma$ contain the following contributions:
\eq{
	\M_V &= \M_{VP} + \M_{ver}^e + \M_{ver}^p + \M_{box},
	\label{eq.VirtualRCAmplitude}\\
	\M_\gamma &= \M_{ISR} + \M_{FSR}.
}
Considering the contributions of the order of $\alpha^3$ we can write:
\ba
	\sigma
	&\sim&
	\brm{\M_B}^2
	+
	2 \,\Re\br{ \M_B^+ \, \M_V }
	+
	\brm{\M_\gamma}^2\nn\\
	&=&
	\brm{\M_B}^2
	\left [ 
		1 + \frac{2 \,\Re\br{ \M_B^+ \, \M_V }}{\brm{\M_B}^2} + \frac{\brm{\M_\gamma}^2}{\brm{\M_B}^2}
	\right ].
	\label{eq.RadiativeCorrectionsDecomposition}
\ea
Therefore cross section with radiative effects of order $\alpha^3$ can be written as:
\eq{
	\sigma
	=
	\sigma_B
	\br{
		1 + \delta_V + \delta_\gamma
	}.
	\label{eq.delta}
}

%----------------------------------------------------
\subsection{Virtual photon contributions}
\label{sec.VirtualRC}
%----------------------------------------------------

The one-loop virtual photon contributions have been calculated in the literature
(see, for example \cite{Ahmadov:2010ak}). We recall all of them for completeness here. 
In this section we use the point-like proton approximation, Eq. (\ref{eq.PointLikeProton}).
The total one-loop virtual corrections include the decomposition   (\ref{eq.VirtualRCAmplitude}) 
of the virtual corrections amplitude:
\begin{enumerate}
	\item
	the boson self energies (or vacuum polarization), i.e., $\M_{VP}$,
	\item
	the corrections to proton ($\M_{ver}^p$) and electron ($\M_{ver}^e$) vertices,
	\item
	the two-photon exchange or box-type contributions, $\M_{box}$.
\end{enumerate}

Below we present all these contributions separately.

%----------------------------------------------------
\subsubsection{Boson self-energy}
%----------------------------------------------------

\begin{figure}
\begin{center}
	\begin{tikzpicture}
		% Left proton.
       \draw[charged_compound] (-1.8,1) -- (-1,0);
       \draw[charged_compound] (-1,0) -- (-1.8,-1);
       % Leptons.
       \draw[fermion] (1.8,-1) -- (1,0);
       \draw[fermion] (1,0) -- (1.8,1);
       % Intermediate particles.
       \draw[photon] (-1,0) -- (-0.4,0);
       \draw[photon] (0.4,0) -- (1,0);
       % Vacuum polarization.
       \draw[line width=1,fill=black, pattern=crosshatch] (0,0) circle (0.4);
       \node at (0,0.7) {$\Pi(s)$};
       % Interaction points.
       \draw[fill=black] (1,0) circle (0.05);
       \draw[fill=black] (-1,0) circle (0.05);
   		\draw[fill=black] (-0.4,0) circle (0.05);
   		\draw[fill=black] (0.4,0) circle (0.05);
   \end{tikzpicture}
	\caption{\label{fig.VP} Vacuum polarization radiative corrections. }
\end{center}
\end{figure}
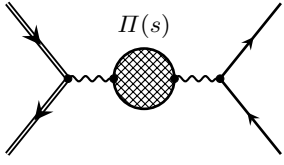

\begin{figure}
\begin{center}
	\begin{tikzpicture}[baseline={([yshift=-3.5mm]current bounding box.center)}]
   		% Intermediate particles.
   		\draw[photon] (-1,0) -- (-0.4,0);
   		\draw[photon] (0.4,0) -- (1,0);
   		% Vacuum polarization.
       \draw[line width=1,fill=black, pattern=crosshatch] (0,0) circle (0.4);
       \node at (0,0.7) {$\Pi(s)$};
   		% Interaction points.
   		\draw[fill=black] (-0.4,0) circle (0.05);
   		\draw[fill=black] (0.4,0) circle (0.05);
   	\end{tikzpicture}
   	=
	\begin{tikzpicture}[baseline={([yshift=-2.5mm]current bounding box.center)}]
   		% Intermediate particles.
   		\draw[photon] (-1,0) -- (-0.4,0);
   		\draw[photon] (0.4,0) -- (1,0);
   		% Vacuum polarization.
   		\draw[line width=1] (0,0) circle (0.4);
   		\draw[fermion] (0.09,0.4) -- (0.095,0.4);
   		\draw[fermion] (-0.09,-0.4) -- (-0.095,-0.4);
   		% Interaction points.
   		\draw[fill=black] (-0.4,0) circle (0.05);
   		\draw[fill=black] (0.4,0) circle (0.05);
   		% Label.
   		\node at (0,0.6) {$e,\mu,\tau$};
   	\end{tikzpicture}
   	+
   	\begin{tikzpicture}[baseline={([yshift=-2.5mm]current bounding box.center)}]
   		% Intermediate particles.
   		\draw[photon] (-1,0) -- (-0.4,0);
   		\draw[photon] (0.4,0) -- (1,0);
   		% Vacuum polarization.
   		\draw[charged_scalar] (0,0) circle (0.4);
   		\draw[charged_scalar] (0.09,0.4) -- (0.095,0.4);
   		\draw[charged_scalar] (-0.09,-0.4) -- (-0.095,-0.4);
   		% Interaction points.
   		\draw[fill=black] (-0.4,0) circle (0.05);
   		\draw[fill=black] (0.4,0) circle (0.05);
   		% Label.
   		\node at (0,0.6) {$\pi$};
   	\end{tikzpicture}
	\caption{\label{fig.BSE} Contributions to the boson self energy operator. }
\end{center}
\end{figure}
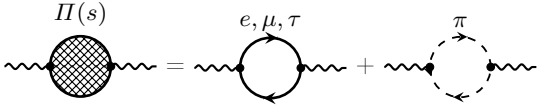

The boson self-energy  contribution corresponds to the diagram shown in Fig.~\ref{fig.VP}
where the vacuum polarization operator $\Pi(s)$ has the following terms 
(see Fig.~\ref{fig.BSE}):
\eq{
	\Pi(s) = \Pi_e(s) + \Pi_\mu(s) + \Pi_\tau(s) + \Pi_{\rm hadr}(s),
}
where
\ba
	\Pi_e(s)   &=& \frac{\alpha}{3\pi} \br{L_e-\frac{5}{3}} - i \frac{\alpha}{3}, \\
	\Pi_\mu(s) &=&-\frac{\alpha}{\pi}\left [ \frac{8}{9}-\frac{\beta_{\mu}^2}{3}-
	\frac{\beta_\mu}{2} \br{1-\frac{\beta_\mu^2}{3}}L_\mu\right ] \nn\\
	&& -i\frac{\alpha}{2} \br{1-\frac{\beta_\mu^2}{3}},
      \label{eq.vacuum}
\ea
where we use the following notations for the logarithms:
\ga{
	L_e \equiv \ln \frac{s}{m^2},
	\qquad
	L_\mu \equiv \ln \frac{1+\beta_\mu}{1-\beta_\mu}.
}
The $\tau$-lepton contribution $\Pi_\tau$ can be found by substitution 
$\Pi_\tau(s) = \Pi_\mu(s)|_{\mu \rightarrow \tau}$.
To evaluate the hadronic contribution we consider the charged pion loop contribution:
\ba
	\Pi_{\rm hadr}(s) &\approx& \Pi_{\pi^+\pi^-}(s)
	=
	\frac{2\alpha}{\pi}\br{
		\frac{1}{12} L_\pi - \frac{2}{3}-2\beta_\pi^2 - i \frac{\beta_{\pi}^3}{12}
	},
\nn\\
   &&
	L_\pi \equiv \ln \frac{1+\beta_\pi}{1-\beta_\pi},
	\ 
	\beta_\pi = \sqrt{1 - \frac{4m_\pi^2}{s}}.
\label{eq.lpi}
\ea

%----------------------------------------------------
\subsubsection{Vertex corrections}
%----------------------------------------------------

\begin{figure}
\begin{center}
	\begin{tikzpicture}
		\draw[charged_compound] (-1.5,1) -- (-1.1,0.5);
		\draw[plain_compound] (-1.1,0.5) -- (-0.7,0);
		\draw[plain_compound] (-0.7,0) -- (-1.1,-0.5);
		\draw[charged_compound] (-1.1,-0.5) -- (-1.5,-1);
		% ã•Ø‚Æ≠Î.
		\draw[fermion] (1.5,-1) -- (0.7,0);
		\draw[fermion] (0.7,0) -- (1.5,1);
		% è‡Æ¨•¶„‚ÆÁ≠Î• Á†·‚®ÊÎ.
		\draw[photon] (-0.7,0) -- (0.7,0);
		\draw[photon] (-1.1,-0.5) to [bend left=40] (-1.1,0.5);
		% íÆÁ™® ¢ß†®¨Æ§•©·‚¢®Ô.
		\draw[fill=black] (0.7,0) circle (0.05);
		\draw[fill=black] (-0.7,0) circle (0.05);
		\draw[fill=black] (-1.1,-0.5) circle (0.05);
		\draw[fill=black] (-1.1,0.5) circle (0.05);
		% Vertex function.
		\draw[dashed] (-1,0) circle (0.65);
		\node at (-2.3,0) {$F_{1,2}^{(2)}(q^2)$};
   \end{tikzpicture}
   \qquad
	\begin{tikzpicture}
		% Proton line.
		\draw[charged_compound] (-1.5,1) -- (-0.7,0);
		\draw[charged_compound] (-0.7,0) -- (-1.5,-1);
		% Electron line.
		\draw[fermion] (1.5,-1) -- (1.1,-0.5);
		\draw[plain] (1.1,-0.5) -- (0.7,0);
		\draw[plain] (0.7,0) -- (1.1,0.5);
		\draw[fermion] (1.1,0.5) -- (1.5,1);
		% Intermediate particles.
		\draw[photon] (-0.7,0) -- (0.7,0);
		\draw[photon] (1.1,-0.5) to [bend right=40] (1.1,0.5);
		% Interaction points.
		\draw[fill=black] (0.7,0) circle (0.05);
		\draw[fill=black] (-0.7,0) circle (0.05);
		\draw[fill=black] (1.1,-0.5) circle (0.05);
		\draw[fill=black] (1.1,0.5) circle (0.05);
		% Vertex function.
		\draw[dashed] (1,0) circle (0.65);
  		\node at (2.3,0) {$F_e^{(2)}(q^2)$};
   \end{tikzpicture}
	\caption{\label{fig.VertexRC} Vertex radiative corrections. }
\end{center}
\end{figure}
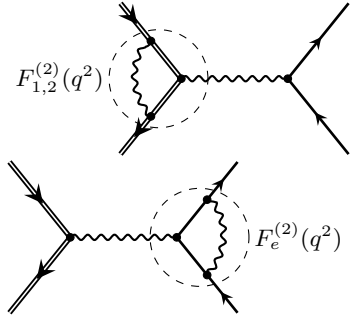

The vertex corrections correspond to the diagrams shown in Fig.~\ref{fig.VertexRC}
where the proton and electron vertices are modified by additional virtual photon exchanges.
This contribution leads to the following modification of proton and electron FFs of the order $\alpha$
with respect to the Born level contributions (\ref{eq.PointLikeProton}):
\eq{
	F_1(q^2) &= 1 + \frac{\alpha}{\pi} F_1^{(2)}(q^2),
	\nn\\
	F_2(q^2) &= \frac{\alpha}{\pi} F_2^{(2)}(q^2),
	\\
	F_e(q^2) &= 1 + \frac{\alpha}{\pi} F_e^{(2)}(q^2),
	\nn	
}
where the functions $F_i^{(2)}$ are given below.
The modification of proton vertex then reads as:
\ba
	\Re \, F_1^{(2)}
	&=&
	\br{ \ln\frac{M}{\lambda}-1 } \br{ 1-\frac{1+\beta^2}{2\beta} L_\beta }+
      \nn\\
	&&
          \frac{1+\beta^2}{2\beta}\left[ 
		\frac{\pi^2}{3} +\Li{2}{\frac{1-\beta}{1+\beta}} - \frac{L^2_\beta}{4} - 
\right  .\\
&&
\left .
     L_\beta\ln\frac{2\beta}{1+\beta}
	\right ] 
	- \frac{L_\beta}{4\beta},
	\Re \, F_2^{(2)}
\nn \\
	&=&
	-\frac{1-\beta^2}{4\beta}L_\beta,
	\ 
	L_\beta \equiv \ln\frac{1+\beta}{1-\beta}.
 \label{eq.eqf}
\ea
This contribution is infrared divergent and is regularized  by introducing
a fictitious photon mass $\lambda$ which cancels with the soft real photon emission contribution
according to the procedure described in \cite{Bloch:1937pw}.

The electron vertex modification leads to
\eq{
	\Re \, F_e^{(2)}
	=
	\br{ \ln\frac{m}{\lambda}-1 } \br{1-L_e}
	-\frac{L_e^2}{4} - \frac{L_e}{4} + \frac{\pi^2}{3}.
}

%----------------------------------------------------
\subsubsection{Box-type corrections}
\label{sec.Box}
%----------------------------------------------------

\begin{figure}
\begin{center}
	\begin{tikzpicture}
		% Proton line.
		\draw[charged_compound] (-1.5,1) -- (-0.7,0.8);
		\draw[charged_compound] (-0.7,0.8) -- (-0.7,-0.8);
		\draw[charged_compound] (-0.7,-0.8) -- (-1.5,-1);
		% Electron line.
		\draw[fermion] (1.5,-1) -- (0.7,-0.8);
		\draw[fermion] (0.7,-0.8) -- (0.7,0.8);
		\draw[fermion] (0.7,0.8) -- (1.5,1);
		% Intermediate particles.
		\draw[photon] (-0.7,0.8) -- (0.7,0.8);
		\draw[photon] (-0.7,-0.8) -- (0.7,-0.8);
		% Interaction points.
		\draw[fill=black] (0.7,0.8) circle (0.05);
		\draw[fill=black] (-0.7,0.8) circle (0.05);
		\draw[fill=black] (0.7,-0.8) circle (0.05);
		\draw[fill=black] (-0.7,-0.8) circle (0.05);
   \end{tikzpicture}
   \qquad
	\begin{tikzpicture}
		% Proton line.
		\draw[charged_compound] (-1.5,1) -- (-0.7,0.8);
		\draw[charged_compound] (-0.7,0.8) -- (-0.7,-0.8);
		\draw[charged_compound] (-0.7,-0.8) -- (-1.5,-1);
		% Electron line.
		\draw[fermion] (1.5,-1) -- (0.7,-0.8);
		\draw[fermion] (0.7,-0.8) -- (0.7,0.8);
		\draw[fermion] (0.7,0.8) -- (1.5,1);
		% Intermediate particles.
		\draw[photon] (-0.7,0.8) -- (0.7,-0.8);
		\draw[photon] (-0.7,-0.8) -- (0.7,0.8);
		% Interaction points.
		\draw[fill=black] (0.7,0.8) circle (0.05);
		\draw[fill=black] (-0.7,0.8) circle (0.05);
		\draw[fill=black] (0.7,-0.8) circle (0.05);
		\draw[fill=black] (-0.7,-0.8) circle (0.05);
   \end{tikzpicture}
	\caption{\label{fig.BoxRC} Box-type radiative corrections. }
\end{center}
\end{figure}
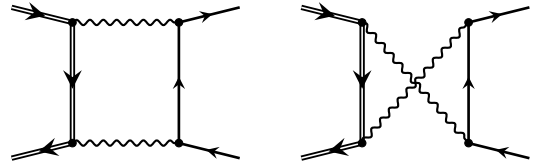

The box-type contributions are illustrated by the diagrams presented in Fig.~\ref{fig.BoxRC}
and implement two virtual photons exchange mechanism.
The box-type contributions are illustrated by the diagrams in Fig.~\ref{fig.BoxRC} and describe the exchange of two virtual photons.
The interference of the box amplitude $\M_{box}$ (\ref{eq.VirtualRCAmplitude})
with the Born amplitude $\M_B$ (\ref{eq.RadiativeCorrectionsDecomposition})
is expressed through the following function:
\ba
	I && (s,t,u) =
	(u-t)
	\nn\\
	&&\left [
		\br{\frac{2M^2}{\beta^2} + t + u} I_{0qp}
		-
		\frac{\pi^2}{6} + \frac{1}{2} L_\beta^2 -
		\frac{1}{\beta^2} L_\beta	\right ]+  \nn\\
		&&
	  (2t+s)\left [\frac{1}{2} L_{ts}^2 - 
	\Li{2}{\frac{-t}{M^2-t}}
	\right ]
	-
	\nn\\
	&&
	(2u+s)\left [\frac{1}{2} L_{us}^2 -
	\Li{2}{\frac{-u}{M^2-u}}\right ]
	+
	\nn \\
	&&
	[ut-M^2(s+M^2)]\left (
		\frac{1}{t} L_{ts} - 
		\frac{1}{u} L_{us} + \frac{u-t}{ut} L_s
	\right ) 
	+
	\nn\\
	&&
	I_0 L_{tu} \br{L_{M\lambda}+L_s},
	\label{eq.BoxFunction}
\ea
where
\ba
	I_{0qp} &=& \frac{1}{s\beta}\Biggl[
		L_s L_\beta - \frac{1}{2} L_\beta^2 - \frac{\pi^2}{6}
		+
		2\Li{2}{\frac{1+\beta}{2}} 
		\nn\\
		&&- 2\Li{2}{\frac{1-\beta}{2}}
		-
		2\Li{2}{\frac{\beta-1}{\beta+1}}
	\Biggr]
\label{eq.ioqp}
\ea
and for the logarithms we use the following notations:
\ba
	L_{ts} &\equiv& \ln\frac{M^2-t}{s},
	\ 
	L_{us} \equiv \ln\frac{M^2-u}{s},
	\ 
	L_s \equiv \ln\frac{s}{M^2},
	\nn\\
	&&L_{tu} \equiv \ln\frac{M^2-t}{M^2-u},
	\ 
	L_{M\lambda} \equiv \ln\frac{M^2}{\lambda^2}.
	\label{eq.eql}
\ea
It is worth to be noted that the coefficient in front of $L_{M\lambda}$ in (\ref{eq.BoxFunction})
\eq{
	I_0 = \frac{2}{s}\br{t^2+u^2-4M^2(t+u)+6M^4} = s\br{2-\beta^2\sin^2\theta}
%\label{eq:23}
}
is proportional to the modulus squared of the Born amplitude.
%----------------------------------------------------
\subsubsection{Total virtual corrections}
%----------------------------------------------------

Summing the Born contribution and all virtual corrections presented above
the differential cross section can be written in the form:
\ba 
	\sigma_{BV}
	&=&
	\frac{\alpha^2}{4 s \beta} \br{2-\beta^2 \sin^2\theta}
	\brm{\frac{1}{1-\Pi(s)}}^2 +
	\frac{\alpha^3}{2 \pi s \beta}\times \nn\\
	&&
	\left \{\left[ 
			(2-\beta^2 \sin^2\theta)
			\Re \left ( F_e^{(2)} + F_1^{(2)}
			+
			2 \, \Re \, F_2^{(2)} 
		\right )\right]\right .
		+
		\nn\\
		&&
		\left . \frac{I(s,t,u)}{s}
	\right \}.
	\label{eq.BornVirt}
\ea
Again we note that quantities $F_1^{(2)}$, $F_e^{(2)}$ and $I(s,t,u)$
contain infrared divergences that are regularized in terms of $\ln\lambda$. Those divergent terms
cancel in the sum with soft photon emission contribution, see below for details.

%----------------------------------------------------
\subsection{Soft real photon emission}
\label{sec.SoftRC}
%----------------------------------------------------

\begin{figure}
 \begin{center}
	\begin{tikzpicture}
        % Proton line.
        \draw[charged_compound] (-1.5,1) -- (-1.1,0.5);
        \draw[plain_compound] (-1.1,0.5) -- (-0.7,0);
        \draw[charged_compound] (-0.7,0) -- (-1.5,-1);
        % Electron line.
        \draw[fermion] (1.5,-1) -- (0.7,0);
        \draw[fermion] (0.7,0) -- (1.5,1);
        % Intermediate particles.
        \draw[photon] (-0.7,0) -- (0.7,0);
        % Bremsstrahlung.
        \draw[photon] (-1.1,0.5) -- (-0.5,1);
        % Interaction points.
        \draw[fill=black] (-1.1,0.5) circle (0.05);
        \draw[fill=black] (0.7,0) circle (0.05);
        \draw[fill=black] (-0.7,0) circle (0.05);
   \end{tikzpicture}
   \qquad
	\begin{tikzpicture}
		% Proton line.
        \draw[charged_compound] (-1.5,1) -- (-0.7,0);
        \draw[plain_compound] (-1.1,-0.5) -- (-0.7,0);
        \draw[charged_compound] (-1.1,-0.5) -- (-1.5,-1);
        % Electron line.
        \draw[fermion] (1.5,-1) -- (0.7,0);
        \draw[fermion] (0.7,0) -- (1.5,1);
        % Intermediate particles.
        \draw[photon] (-0.7,0) -- (0.7,0);
        % Bremsstrahlung.
        \draw[photon] (-1.1,-0.5) -- (-0.5,-1);
        % Interaction points.
        \draw[fill=black] (-1.1,-0.5) circle (0.05);
        \draw[fill=black] (0.7,0) circle (0.05);
        \draw[fill=black] (-0.7,0) circle (0.05);
   \end{tikzpicture}
   \qquad
	\begin{tikzpicture}
		% Proton line.
        \draw[charged_compound] (-1.5,1) -- (-0.7,0);
        \draw[charged_compound] (-0.7,0) -- (-1.5,-1);
        % Electron line.
        \draw[fermion] (1.5,-1) -- (0.7,0);
        \draw[plain] (0.7,0) -- (1.1,0.5);
        \draw[fermion] (1.1,0.5) -- (1.5,1);
        % Intermediate particles.
        \draw[photon] (-0.7,0) -- (0.7,0);
        % Bremsstrahlung.
        \draw[photon] (1.1,0.5) -- (1.5,0);
        % Interaction points.
        \draw[fill=black] (1.1,0.5) circle (0.05);
        \draw[fill=black] (0.7,0) circle (0.05);
        \draw[fill=black] (-0.7,0) circle (0.05);
   \end{tikzpicture}
   \qquad
	\begin{tikzpicture}
		% Proton line.
        \draw[charged_compound] (-1.5,1) -- (-0.7,0);
        \draw[charged_compound] (-0.7,0) -- (-1.5,-1);
        % Electron line.
        \draw[fermion] (0.7,0) -- (1.5,1);
        \draw[plain] (1.1,-0.5) -- (0.7,0);
        \draw[fermion] (1.5,-1) -- (1.1,-0.5);
        % Intermediate particles.
        \draw[photon] (-0.7,0) -- (0.7,0);
        % Bremsstrahlung.
        \draw[photon] (1.1,-0.5) -- (1.5,0);
        % Interaction points.
        \draw[fill=black] (1.1,-0.5) circle (0.05);
        \draw[fill=black] (0.7,0) circle (0.05);
        \draw[fill=black] (-0.7,0) circle (0.05);
   \end{tikzpicture}
	\caption{\label{fig.BremRC} Real photon emission (Bremsstrahlung). }
\end{center}
\end{figure}
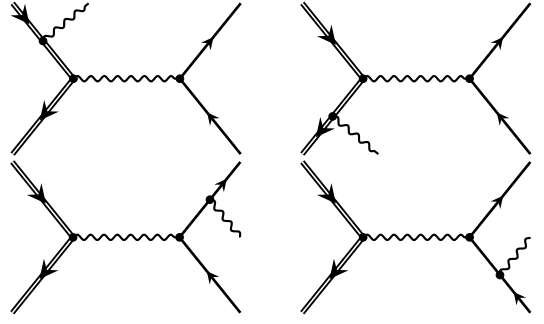

The real photon emission process
\eq{
	p(p_1)+\bar{p}(p_2) \to e^+(k_1)+e^-(k_2)+\gamma(k),
	\label{eq.Bremsstrahlung}
}
is illustrated by the diagrams of  Fig.~\ref{fig.BremRC}). It has to be taken into account since in any detection there is an 
energy threshold $\Delta E \ll \sqrt{s}$ below which the emitted  photons are not detected. Such  
events can not be considered as elastic events. This is the reason why pure elastic events are not
physical and one of the consequence of this fact is that the virtual corrections discussed in the previous section have infrared
divergence. Nevertheless the measured quantities are finite and with a definite physical meaning. In
case of radiatively corrected cross section (at the next to leading order in $\alpha$), the minimal
experimentally measurable set of contributions contains Born terms, virtual corrections and emission of one real soft photon (i.e., the photon energy $\omega$ is smaller than the experimental threshold,
$\omega < \Delta E$). Again the  infrared divergence of the soft photon emission is regularized with a fictitious photon mass
$\lambda$ (i.e., $\omega > \lambda$). In this approach the sum of all the contributions mentioned above is finite
and does not depend on the photon mass $\lambda$. This procedure is the standard implementation of Bloch and Nordsieck idea
\cite{Bloch:1937pw} of infrared divergencies cancellation.

Thus, the soft photon emission is evaluated in the soft regime where the photon c.m.s. energy, $\omega$, is
constrained by $\lambda < \omega < \Delta E \ll E =\sqrt{s}/2$. All numerical applications in this paper are done for
$\Delta E/E = 0.01$.

The charge even contribution to soft photon emission (the sum of the ISR and FSR contributions) was
calculated in \cite{Ahmadov:2010ak}:
\ba
	\frac{d\sigma^{\rm soft}_{\rm even}}{d\sigma_B}
	&=&
	\frac{\alpha}{\pi}
	\left[
		-2 \br{\ln \frac{2\Delta E}{\lambda} - \frac{1}{2\beta} L_{\beta}}
		-2\ln\br{\frac{\Delta E}{E}\frac{m}{\lambda}}
	\right .
	+
	\nn \\
	&& 2
		\frac{1+\beta^2}{2\beta}\br{\ln\frac{2 \Delta E}{\lambda} L_{\beta} - \frac{1}{4} L_{\beta}^2 + \Phi(\beta)}+
		\nn\\
		&&\left . 
		2 \left ( \ln \frac{2\Delta E}{\lambda} L_e -\frac{1}{4} L_e^2-\frac{\pi^2}{6}\right )
	\right],
	\label{eq.SoftEven}
\ea
where the function $\Phi(\beta)$ is:
\ba
	\Phi(\beta) &=&
	\frac{\pi^2}{12} + L_\beta\ln\frac{1+\beta}{2\beta}
	+
	\ln\frac{2}{1+\beta}\log\br{1-\beta}
	+
	\nn\\
	&&
	\frac{1}{2}\ln^2\br{1+\beta}
	-
	\frac{1}{2}\ln^2 2
	-
\nn\\
	&&
	\Li{2}{\beta}
	+
	\Li{2}{-\beta}-
	\Li{2}{\frac{1-\beta}{2}},
	\label{eq:eqA4}
\ea
and satisfies the relation $\Phi(1)=-\pi^2/6$.
The expression (\ref{eq.SoftEven}) contains the sum of ISR and FSR terms and assumes a kinematic regime
where $s, -t, -u \gg M^2, m^2$ that may not hold for PANDA kinematics, in particular at very forward or very backward
photon angles.

This contribution was also calculated in \cite{VandeWiele:2012nb}, where the term
$\frac{1}{\beta} \ln\br{\frac{1+\beta}{1-\beta}}$ was written in ultrarelativistic form, i.e., $\ln\br{s/M^2}$. This
approximation holds when $s \gg M^2$, which is not the case for PANDA rather small energies
($2.25 \geq \sqrt{s} \geq 5.56~\GeV$). Numerically, however, the difference due to this approximation is not large in
the kinematics investigated here.

Here we recalculate these charge even contributions separately.
The soft photon emission from the proton line (initial state radiation, ISR) can be written in a more symmetric form:
\ba
	\frac{d\sigma^{\rm soft}_{\rm ISR}}{d\sigma_B}
	&=&
	\frac{\alpha}{\pi}
	\left \{
		\br{
			\frac{1+\beta^2}{\beta} L_\beta - 2
		}
		\ln\br{\frac{2\Delta E}{\lambda}}
		+
		\frac{1}{\beta} L_\beta
		+\right .
		\nn\\
		&&\left .
		\frac{ 1+ \beta^2 }{2 \beta}
		\left [
			\Li{2}{\frac{2\beta}{\beta-1}}
			-
			\Li{2}{\frac{2\beta}{\beta+1}}
	\right ]\right \},
	\label{eq.SoftISR}
\ea
that coincides with Eq. (29) in \cite{Berends:1973tz}.
The soft photon emission from the electron line (final state radiation, FSR)
can be obtained from the expression above with the following replacement:
\eq{
	\sigma^{\rm soft}_{\rm FSR} = \left.\sigma^{\rm soft}_{\rm ISR}\frac{}{}\right|_{p \to e}.
	\label{eq.SoftFSR}
}
This result for FSR agrees with formula (29) from \cite{Berends:1973tz}
and it is in a good agreement with equation (36) in \cite{VandeWiele:2012nb}
because the approximation $\frac{1}{\beta_e} \log\frac{1+\beta_e}{1-\beta_e} \approx \log\br{s/m^2}$
holds for PANDA energies with very good accuracy.

Charge odd contributions arise from the interference of ISR and FSR terms of the amplitude.
Let us note here that this contribution was calculated in Ref. \cite{VandeWiele:2012nb},
 Eqs (37) and (38), but this result 
is not symmetric with respect to the masses of proton and electron.
This contribution is recalculated here following the formalism of \cite{Berends:1973fd}, leading to 
 a symmetric result (both under the particle masses and under the Mandelstam invariants interchange):
\ba
	\frac{d\sigma^{\rm soft}_{\rm odd}}{d\sigma_B}
	&=
	-\frac{ \alpha}{2\pi^2}
	\Bigl [& \br{m^2+M^2-t} R(s,t) - 
	\nn\\
	&&
	\br{m^2+M^2-u} R(s,u)\Bigr ] ,
	\label{eq.SoftOdd}
\ea
where the function $R$ is defined in formula (A.11) from \cite{Berends:1973fd} and has the form:
\ba
	R(s,t) &=& 2\pi\br{2 A(s,t) \ln\frac{2\Delta E}{\lambda} + C(s,t)},
	\label{eq:eqR}
	\\
	A(s,t) &=& \frac{1}{\sqrt{\lambda(t,m^2,M^2)}} \times
\nn\\
&& 
        \ln \brm{
		\frac{t-m^2-M^2-\sqrt{\lambda(t,m^2,M^2)}} {t-m^2-M^2 + \sqrt{\lambda(t,m^2,M^2)}}
	},
	\nn\\
	C(s,t) &=& \frac{1}{\sqrt{\lambda(t,m^2,M^2)}} \sum_{i,j=1}^4
	\epsilon_i\, \delta_j \, U_{ij}(\eta_0,\eta_1,y_i,y_j),
	\label{eq.C}
\nn
\ea
where $$\lambda(x,y,z)=x^2+y^2+z^2-2xy-2xz-2yz$$,  
$$\epsilon_i=(+1,-1,-1,+1), \ \delta_j=(-1,-1,+1,+1)$$,
and
\eq{
	\eta_0 &= \sqrt{1-m^2/E^2},
	\nn\\
	\eta_1 &= \sqrt {1-M^2/E^2}+\sqrt{-t}/E,
	\nn\\
	y_i &= \delta_i - \frac{t+m^2-M^2 + \epsilon_i \delta_i \sqrt{\lambda(t,m^2,M^2)} }{2E\sqrt{-t} }.
	\nn
}
The functions $U_{ij}$ in Eq.~(\ref{eq.C}) have the form:
\ba
	U_{ij} &=& \Re \br{ \Li{2}{\frac{\eta_0-y_i}{y_j-y_i}} - \Li{2}{\frac{\eta_1-y_i}{y_j-y_i}} }
	+
	\nn\\
	&&
	\ln\brm{y_i-y_j} \ln\brm{\frac{\eta_1-y_i}{\eta_0-y_i}},\qquad\mbox{for}\quad i \ne j,
	\nn\\
	U_{ii} &=& \frac{1}{2}\ln^2\brm{\eta_1-y_i} - \frac{1}{2}\ln^2\brm{\eta_0-y_i}.
	\nn
\ea
%----------------------------------------------------
\subsection{Hard real photon emission}
\label{sec.HardRC}
%----------------------------------------------------

In some experimental setup it is not possible to separate elastic events (including
soft photon emission) and inelastic events, when an additional  hard photon is emitted. Even if one can detect such events, the   energy threshold of the detection may be  too high and the soft photon approximation is no more valid.
In these situations one needs to take into account precisely  hard photon emission, without approximations as far as it is possible.

The calculation of hard photon emission requires to take into account
the proton structure, what is not necessary in the case of soft photon approximation considered in Section~\ref{sec.SoftRC}.
Note that the  box-type diagrams considered in Section~\ref{sec.Box} suffer of the same
difficulties and they are solved in a similar way as described in this section.

A simple  way to take the proton structure into account consists in introducing proton
form factors in the proton-photon vertices in the diagrams of Fig.~\ref{fig.BremRC}.
However, this leads to gauge invariance violation and requires a specific treatment to restore this invariance. 
Moreover,  in the first two diagrams of Fig.~\ref{fig.BremRC}, the intermediate proton propagator is 
off-mass-shell and the usual form factors
$F_{1,2}\br{q^2}$ cannot be used in this context, since they are defined for the case when both proton
lines in the vertex are on-mass-shell. 

In such case, one cannot just use vertices with off-mass-shell external legs (extrapolated in some model-dependent way).
It is also necessary to take into account any possible "excited proton propagator". For example, in a 
meson--baryon approach,  other diagrams must be taken into account: $\Delta$-resonance intermediate state, extra pion exchange connected to the
proton line, two pion exchange, three pion exchange etc. A consistent calculation would be realistic and gauge invariant but there is no consensus  about how to proceed.

Therefore, at the light of these arguments,  we estimate the hard photon emission contribution in the following way:
\begin{enumerate}

	\item
		We calculate the Born $\sigma_B^0$ and hard photon emission $\sigma_\gamma$ cross sections using the 
		point-like proton approximation (\ref{eq.PointLikeProton})
		and estimate the correction $\delta_\gamma$ from (\ref{eq.delta}) to the ratio of these two cross sections:
		\eq{
			\delta_\gamma = \frac{\sigma_\gamma}{\sigma_B^0}.
		}
		
	\item
		The corrected cross section is then estimated using Eq. (\ref{eq.delta}), where
		Born cross section $\sigma_B$ is calculated from Eq. (\ref{eq.BornWithFF}), including 
		proton form factors.

\end{enumerate}
This approach allows us to keep the gauge invariance and the consistency of our calculation
 obtaining a realistic value of the corrected cross section which, in case of point-like proton, would be
much larger than the experimental values.

The previous calculations of hard photon emission for the reaction of interest contain also approximations. 
In Ref. \cite{Ahmadov:2010ak} the calculation of hard photon emission is performed 
under a similar assumptions of point-like proton, but  in ultra relativistic approximation
when $s \gg M^2$, which is not the case for PANDA. This calculation is exact 
for the reaction $e^+ e^- \to \mu^+ \mu^-$ and can be compared to the results of \cite{Kuraev:1977rp}.
The authors of Ref. \cite{VandeWiele:2012nb} attempted to restore the proton mass dependence and the role of the proton structure.
 For the calculation of  hard photon emission, the proton structure is present in the Pauli form factor, related to the anomalous magnetic moment of the proton (see Eq. (C.2) in \cite{VandeWiele:2012nb}). This approach is inconsistent, as virtual corrections (modified vertices and box-type diagrams) are evaluated in point-like approximation and neither form factors nor dynamical structure of the proton are included.

The aim of this section is to calculate the hard photon contribution and give expressions that can be
applied to the conditions of acceptance and kinematics of the PANDA detector. The calculation follows these steps:
\begin{enumerate}

	\item
		
		The cross section of the radiative process (\ref{eq.Bremsstrahlung}) is written as the squared amplitude
		of the real photon emission process, setting the following condition on the energy of the
		emitted photon in c.m.s of the initial particles: $\omega > \Delta E$.
		This condition characterizes the definition of hard photon.
	
	\item
		The amplitudes for ISR (fist two diagrams in Fig.~\ref{fig.BremRC}) and for FSR 
		(the last two diagrams in Fig.~\ref{fig.BremRC}) are written and coherently summed,
		giving a term for ISR, a term for FSR and an interference term (INT).

	\item
		Radiative invariants are  chosen and expressed through the observable quantities.

	\item
		The phase space and the Gram determinant are expressed in terms of these invariants.

	\item
		A factor $\rm\Theta_P$ (under the integral) is defined in order to implement the necessary 
		experimental cuts including the conditions of the PANDA detector.

	\item
		The numerical integration is done and the independence of the result of the choice of $\Delta E$
		is verified.
		
\end{enumerate}

The differential hard photon cross section for the process (\ref{eq.Bremsstrahlung}) is:
\ba
	d\sigma_\gamma^{\rm hard}
	&=&
	\frac{\alpha^3}{\pi^2 s} \int (\M_{\rm ISR}+\M_{\rm FSR})(\M_{\rm ISR}+\M_{\rm FSR})^+
	\times
	\nn\\
	&&
			\ \ \ {\rm \Theta_\omega} \cdot  {\rm\Theta_P}  \cdot d\Phi_3,
	\nn
\ea
where $\theta_\omega = \theta\br{\omega - \Delta E}$ is the factor that limits the energy region for the
hard photon and
\eq{
	d\Phi_3 = \delta\br{p_1+p_2-k_1-k_2-k} \frac{d \vv{k_1}}{2 \varepsilon_1} \frac{d \vv{k_2}}{2 \varepsilon_2}
	\frac{d \vv{k}}{2 \omega}
}
is the full phase space of the reaction. Here $\varepsilon_{1,2}$ and $\vv{k_{1,2}}$ are the c.m.s energies
and 3-momenta of the final electron and positron, 
respectively, and
$\omega$ and $\vv{k}$ are the energy and 3-momentum of the emitted hard photon.

The matrix elements (amplitudes) corresponding to ISR and FSR (see Fig.~\ref{fig.BremRC})
take the form:
\ba
	\M_{\rm ISR}
	&=&
	\frac{i e_\rho(k) Q_p^2 Q_e}{\br{k_1+k_2}^2}
	\bar{u}(-p_2) \Biggl[
		\Gamma^\mu(q_1^2) \frac{-\dd{p_1} + \dd{k} - M}{2 \br{p_1 k}}\times 
			\nn \\ 
			&&
		 \Gamma^\rho(k^2)
		+\Gamma^\rho(k^2) \frac{\dd{p_2} - \dd{k} - M}{2 \br{p_2 k}} \Gamma^\mu(q_1^2)
	\Biggr ] 
		\nn \\ 
			&&
	u(p_1)
	\bar{u}(k_2) \gamma_\mu u(-k_1),
	\label{eq.MIsr}\\ 
	\M_{\rm FSR}
	&=&
	\frac{i e_\rho(k) Q_p Q_e^2}{\br{p_1+p_2}^2}
	\bar{u}(k_2)  \Biggl[
		\gamma^\mu \frac{-\dd{k_1} - \dd{k} + m}{2 \br{k_1 k}} \gamma^\rho
		+
			\nn \\ &&
		\gamma^\rho \frac{\dd{k_2} + \dd{k} + m}{2 \br{k_2 k}} \gamma^\mu
	\Biggr ]  u(-k_1)
	\bar{u}(-p_2) \Gamma_\mu(q_1^2) u(p_1),
	\nn
\ea 
where $Q_f$ is electric charge of the particle $f$ in units of proton charge,
$q_1$ is the momentum transfer after hard photon  emission, i.e., $q_1 = q - k$,
and the usual expression for proton vertex is
\eq{
	\Gamma^\mu(q^2) =  F_1(q^2)\gamma^\mu + \frac{F_2(q^2)}{4M}\brs{\gamma^\mu,\dd{q}}.
}
The form factors $F_{1,2}(q^2)$ are taken in the point-like approximation (\ref{eq.PointLikeProton}).
Squaring the amplitude one finds:
\ba 
&&	(\M_{\rm ISR}+\M_{\rm FSR})(\M_{\rm ISR}+\M_{\rm FSR})^+
	\nn\\
&&	=
	R_{\rm ISR}+R_{\rm FSR}+R_{\rm INT} = \sum_i R_i.
\label{eq.eqR}
\ea

The ISR term has the form:
\ba
	R_{\rm ISR}
	&=&
	- \frac{Q_p^4 Q_e^2}{\br{k_1+k_2}^4} \frac{1}{4}
	\Sp\left[
		\left (
			\Gamma^\mu(q_1^2) \frac{-\dd{p_1} + \dd{k} - M}{2 \br{p_1 k}} \Gamma^\rho(k^2)
			+
			\right . \right . 
			\nn\\
&&
\left.
	\Gamma^\rho(k^2) \frac{\dd{p_2} - \dd{k} - M}{2 \br{p_2 k}} \Gamma^\mu(q_1^2)
				\right ) 
	\br{\dd{p_1} - M}
	 \times\nn\\
	&&
		\left (
			\Gamma_\rho(k^2) \frac{-\dd{p_1} + \dd{k} - M}{2 \br{p_1 k}} \Gamma^\nu(q_1^2)
			+	\right . 
	\nn\\
	&&		
	\left. 		\left. 		
			\Gamma^\nu(q_1^2) \frac{\dd{p_2} - \dd{k} - M}{2 \br{p_2 k}} \Gamma_\rho(k^2)
			\right )
	\br{\dd{p_2}+M}
	\right] \times
	\nn\\
	&& \Sp[  \gamma_\mu (\dd{k_1} - m) \gamma_\nu (\dd{k_2} +m) ].
\nn\ea

\ba
	R_{\rm FSR}
	&=&
	- \frac{Q_p^2 Q_e^4}{\br{p_1+p_2}^4} \frac{1}{4}
	\Sp\left[
		\left (
			\gamma^\mu \frac{-\dd{k_1} - \dd{k} + m}{2 \br{k_1 k}} \gamma^\rho +
	\right . 	\right . 		
	\nn\\
	&&
	\left .	
			\gamma^\rho \frac{\dd{k_2} + \dd{k} + m}{2 \br{k_2 k}} \gamma^\mu
		\right )
	\br{\dd{k_1} - m}
	\times\nn\\
	&&
		\left(
			\gamma_\rho \frac{-\dd{k_1} - \dd{k} + m}{2 \br{k_1 k}} \gamma^\nu +
			\right .
				\nn\\
	&&
	\left .	\left .
			\gamma^\nu \frac{\dd{k_2} + \dd{k} + m}{2 \br{k_2 k}} \gamma_\rho
		\right ) 
	\br{\dd{k_2}+m}
	\right ] \times
	\nn\\
	&&
	\Sp\left [  \Gamma_\mu(q_1^2) (\dd{p_1} - M) \Gamma_\nu(q_1^2) (\dd{p_2} +M) 
   \right ].
\ea

The INT term is:
\ba
	R_{\rm INT}
	&=&
	- 2\frac{Q_p^3 Q_e^3}{\br{k_1+k_2}^2 \br{p_1+p_2}^2} \times
	\nn\\
	&&
	\frac{1}{4}
	\Sp\left[
		\left (
			\Gamma^\mu(q_1^2) \frac{-\dd{p_1} + \dd{k} - M}{2 \br{p_1 k}} \Gamma^\rho(k^2)
			\right .\right .
			+
			\nn\\
			&&
			\left .
			\ \Gamma^\rho(k^2) \frac{\dd{p_2} - \dd{k} - M}{2 \br{p_2 k}} \Gamma^\mu(q_1^2)
		\right )
          \times 
	\nn\\
	&&	
		(\dd{p_1} - M) \Gamma_\nu(q_1^2) (\dd{p_2} + M)
\Biggr ]\times
	\nn \\
	&&
	\Sp\left[ 
		\gamma_\mu (\dd{k_1} - m) \left (
			\gamma_\rho \frac{-\dd{k_1} - \dd{k} + m}{2 \br{k_1 k}} \gamma^\nu +
			\right .\right .
		\nn \\
	&&	
	\left .\left .
			\ \gamma^\nu \frac{\dd{k_2} + \dd{k} + m}{2 \br{k_2 k}} \gamma_\rho
		\right )
		(\dd{k_2} + m)
	\right ].
	\nn
\ea

%----------------------------------------------------
\subsubsection{Invariant phase space parameterization}
\label{sec.InvariantPhaseSpace}
%----------------------------------------------------

\begin{table}
\caption{ Nonradiative and radiative Lorentz invariants. }
\label{table.Invariants}
\begin{center}
\begin{tabular}{|c|c|c|}
\hline
  \multicolumn{1}{|c|}{Lorentz-invariant} 
& \multicolumn{1}{ c|}{ Nonradiative } 
& \multicolumn{1}{ c|}{Radiative} \\
\hline 
$s= (p_1+p_2)^2 $ &  Eq.~(\ref{eq.s}) & Eq.~(\ref{eq.s})  \\
$t= (p_2-k_2)^2 $ &  Eq.~(\ref{eq.t}) & Eq.~(\ref{eq.t-rad}) \\
$u= (p_1-k_2)^2 $ &  Eq.~(\ref{eq.u}) & Eq.~(\ref{eq.u-rad})  \\
\hline 
$z_1= 2 \br{p_1 k} $ & 0 & $z+v-v_1$  \\
$v_1= 2 \br{p_2 k} $ & 0 & independent  \\
$z  = 2 \br{k_1 k} $ & 0 & independent  \\
$v  = 2 \br{k_2 k} $ & 0 & independent  \\
\hline
$s_1= (k_1+k_2)^2$ & $s$ & $s-z-v$  \\
$t_1= (p_1-k_1)^2$ & $t$ & $t+v-v_1$  \\
$u_1= (p_2-k_1)^2$ & $u$ & $u+v_1-z$  \\
\hline
\end{tabular}
\end{center}
\end{table}

First, the phase space $d\Phi_3$ is parametrized in terms of invariants (see Table.~\ref{table.Invariants}
for definitions).
It should be noted that only the invariant $s$ does not change  in nonradiative and radiative cases,
all other Mandelstam invariants (\ref{eq.s})-(\ref{eq.u}) are modified in the radiative case.
The invariant $t$ is related to the electron emission angle $\cos\theta$ in the c.m.s by
(see Ref.~\cite{Aleksejevs:2016tjd}):
\eq{
	t = \frac{1}{2} \br{ 2M^2 + 2m^2 - s + z + \beta \cos\theta \sqrt{(s-z)^2-4m^2 s} }.
	\label{eq.t-rad}
}
Then, the radiative $u$ is:
\eq{
	u = \frac{1}{2} \br{ 2M^2 + 2m^2 - s + z - \beta \cos\theta \sqrt{(s-z)^2-4m^2 s} }.
	\label{eq.u-rad}
}
The analogue of the relation (\ref{eq.stu}) in the radiative case is:
\eq{
	s + t + u = z + 2m^2 + 2M^2. 
	\label{eq.stuz}
}
Then real photon energy has the form:
\eq{
	\omega = \frac{v + z}{2\sqrt{s}},
}
and the hard photon cut factor then is
\be
	 {\rm\Theta}_\omega \equiv {\rm\Theta}\br{\omega-\Delta E} = {\rm\Theta}\br{ \frac{v+z}{2\sqrt{s}} - \Delta E }.
	\label{eq.hard-cut}
\ee
The dependence of the differential cross section on the emission angle can be given now in terms
of invariants. For this aim, the phase space $d\Phi_3$ is expressed in form of the standard Gram determinant
$\Delta_4$ \cite{Byckling:1973}:
\eq{
	d\Phi_3 = \frac{\pi}{16 \sqrt{\lambda(s,M^2,M^2)}} \frac{dt \, dv \, dz \, dv_1}{\sqrt{-\Delta_4}}.
%	\label{G3a}
}
Then we can write:
\eq{
	d\sigma_\gamma^{\rm hard}
	=
	\frac{\alpha^3}{8\pi s}
	\int \frac{dv \, dz \, dv_1}{\sqrt{R}} \frac{s-z}{s}
	\sum_i R_i \cdot \theta(R) \cdot  {\rm\Theta_\omega} \cdot  {\rm\Theta_P},
%	\label{dif_R}
}
where the Gram determinant $\Delta_4 = -R/16$ is expressed through 
the determinant of a matrix $G$, that is composed of the 
scalar products of the four 4-momenta:
\eq{
	R = - {\rm det}
	\br{
	\begin{matrix}
		2 \br{p_1p_1} & 2 \br{p_2p_1} & 2 \br{k_1p_1} & 2 \br{k_2p_1} \\
		2 \br{p_1p_2} & 2 \br{p_2p_2} & 2 \br{k_1p_2} & 2 \br{k_2p_2} \\
		2 \br{p_1k_1} & 2 \br{p_2k_1} & 2 \br{k_1k_1} & 2 \br{k_2k_1} \\
		2 \br{p_1k_2} & 2 \br{p_2k_2} & 2 \br{k_1k_2} & 2 \br{k_2k_2}
	\end{matrix}
	}.
}
The acceptance conditions for the PANDA detector have to be implemented. Minimal requirements,
for testing the procedure can be considered:
\be
	 {\rm\Theta_P}\equiv  {\rm\Theta}\br{\omega_{\rm max} - \omega} =  {\rm\Theta}\br{\omega_{\rm max} - \frac{v+z}{2\sqrt{s}} },
\ee 
where $\omega_{\rm max}$ is the maximum energy of the Bremsstrahlung photon (in c.m.s).
The $\theta_P$ factor in general takes the form:
\ba
	 {\rm\Theta_P}
	&\equiv&
	 {\rm\Theta}\br{E_{e^-}^{\rm lab} - E_{e^-}^{\rm min}}  {\rm\Theta}\br{E_{e^-}^{\rm max} - E_{e^-}^{\rm lab}}
	 {\rm\Theta}\br{\theta_{e^-}^{\rm lab} - \theta_{e^-}^{\rm min}} \times 
	\nn\\
	&&
	 {\rm\Theta}\br{\theta_{e^-}^{\rm max} - \theta_{e^-}^{\rm lab}},
	\label{eq.Cuts}
\ea
where the quantities $E_{e^-}^{\rm min}$, $E_{e^-}^{\rm max}$, $\theta_{e^-}^{\rm min}$, $\theta_{e^-}^{\rm max}$
and similar ones for the $e^+$ are given by the experimental conditions.
The quantities $E_{e^-}^{\rm lab}$, $\theta_{e^-}^{\rm lab}$ are then expressed via radiative invariants
\ga{
	E^{\rm lab}_{e^-} = \frac{M^2+m^2-u}{2M},
	\ 
	E^{\rm lab}_{e^+} = \frac{M^2+m^2-t_1}{2M}, \\
	\cos\theta^{\rm lab}_{e^-} = \frac{M(t-M^2-m^2)+E(M^2+m^2-u)}{P\sqrt{(M^2+m^2-u)^2-4m^2M^2}}, \\
	\cos\theta^{\rm lab}_{e^+} = \frac{M(u_1-M^2-m^2)+E(M^2+m^2-t_1)}{P\sqrt{(M^2+m^2-t_1)^2-4m^2M^2}}.
%	\label{e-and-a}
}
and have to be substituted in (\ref{eq.Cuts}) to perform the integratation according to the PANDA detector acceptance.

%----------------------------------------------------
\subsubsection{Physical phase space parameterization}
\label{sec.PhysicalPhaseSpace}
%----------------------------------------------------

\begin{figure}
\begin{center}
    \begin{tikzpicture}
        % é·®.
        \draw[-latex] (0,0)--(0,3);
        \node at (0.2,2.9) {$z$};
        \draw[-latex] (0,0)--(2,0);
        \node at (1.8,0.2) {$y$};
        \draw[-latex] (0,0)--(-1.5,-1);
        \node at (-1.3,-0.65) {$x$};
        % ç†Á†´Æ ™ÆÆ‡§®≠†‚.
        \draw[fill=black] (0,0) circle (0.03);
        % ì£´Î.
	    \draw (0,0)  pic[-latex]{carc=90:35:0.4};
        \node at (0.3,0.57) {$\theta_5$};
	    \draw (0,0)  pic[-latex]{carc=-145:-20:0.2};
        \node at (0,-0.35) {$\varphi_5$};
	    \draw (0,0)  pic[-latex]{carc=90:126:0.5};
        \node at (-0.2,0.7) {$\theta$};
        % à¨Ø„´Ï·Î.
        \draw[-latex,line width=1pt] (0,0)--(0,2.5);
        \node at (0.3,2.2) {$\vec{p}_2$};
        \draw[-latex,line width=1pt] (0,0)--(-1,1.3);
        \node at (-1,0.9) {$\vec{k}_2$};
        \draw[-latex,line width=1pt] (0,0)--(1.3,0.8);
        \node at (1.1,0.45) {$\vec{p}_5$};
        \draw[-latex,line width=1pt] (-1,1.32)--(1.3,0.82);
        \node at (0.5,1.3) {$\vec{k}_1$};
        % è‡Æ•™Ê®®.
        \draw[dashed] (1.3,0.8)--(1.3,-0.5);
        \draw[dashed] (1.3,-0.5)--(1.8,0);
        \draw[dashed] (1.3,-0.5)--(-0.7,-0.5);
        \draw[dashed] (1.3,-0.5)--(0,0);
    \end{tikzpicture}
\caption{\label{fig.Angles} Vector configuration in c.m.s.}
\end{center}
\end{figure}
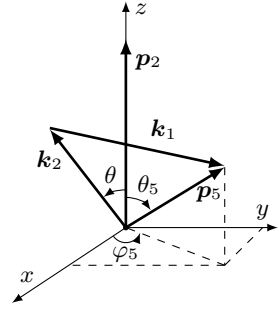

In this section we parametrize phase space $d\Phi_3$ in terms of physical variables which are convenient for
an experimental setup. We follow the $G/N$-method \cite{Zykunov:2017gyf}
(which does not use separation of soft and hard photon energies at all) which uses infinitesimal parameter $\lambda$ as 
a photon mass. The we get the cross section for Bremsstrahlung process in the form:
\ba
	\frac{d\sigma_\gamma^{\rm hard}}{d\cos\theta}
&	=&
	\frac{\alpha^3}{4\pi s} 
	\int\limits_0^{\brm{\vv{k}}_{\rm max}} d\brm{\vv{k}} \frac{\brm{\vv{k}}^2 }{\omega} 
	\int\limits_0^\pi d\theta_k \sin\theta_k
	\int\limits_0^{2\pi} d\varphi_k
	\nn\\
	&&
	\frac{\brm{\vv{k_2}}}{\varepsilon_1 \, g(\varepsilon_2)}
	\sum_i R_i \,  {\rm\Theta_P},
%	\label{dif-brems2}
\ea
where $\brm{\vv{k}}_{\rm max}$ is the maximum 3-momentum (in modulus) of the emitted photon and
$\varepsilon_1$ and $\varepsilon_2$ are the energies of final electron and positron in c.m.s and
\eq{
	g(x)
	&=
	1 + \frac{x \br{1-\brm{\vv{k}} A \br{x^2-m^2}^{-1/2}} }{\sqrt{x^2 - 2\brm{\vv{k}} A \sqrt{x^2-m^2}+\brm{\vv{k}}^2}},
	\\
	A
	&\equiv
	\cos(\widehat{ \vv{p_5},\vv{k_2} }) = \sin\theta \sin\theta_5 \cos\varphi_5 + \cos\theta \cos\theta_5,
%	\label{AN}
}
where the notations of vectors and angles  are illustrated in Fig.~\ref{fig.Angles}.
For convenience we introduce the  auxiliary vector $\vv{p_5}$ which
is defined as $\vv{p_5} \equiv -\vv{k}$, i.e., $\theta_k = \pi-\theta_5$, $\varphi_k = \pi+\varphi_5$
and the components of the 3-vectors  are:
\eq{
	\vv{k_1} &= \vv{p_5} - \vv{k_2}, \  \varepsilon_1 = \sqrt{\brm{\vv{k_1}}^2 + m^2},
	\nn\\
	\vv{k_2} &= \br{ \brm{\vv{k_2}} \sin\theta, 0, \brm{\vv{k_2}}\cos\theta },
	\nn\\
	\vv{p_5} &= \br{ \brm{\vv{k}} \sin\theta_5 \cos\varphi_5, \brm{\vv{k}} \sin\theta_5 \sin\varphi_5, \brm{\vv{k}}\cos\theta_5 },
	\nn
%	\label{vecs}
}
The energy of the positron $\varepsilon_2$ is fixed by energy conservation:
\eq{
	\varepsilon_2 =
	\begin{cases}
		\varepsilon_2^-,\ \mbox{if}\ A>0,
		\\
		\varepsilon_2^+,\ \mbox{if}\ A<0,
	\end{cases}
	\ 
	\varepsilon_2^\pm = \frac{BC \pm \sqrt{C^2+m^2(1-B^2)}}{1-B^2},
}
where
\ga{
	B = \frac{\sqrt{s}-\omega}{A \brm{\vv{k}}},
	\qquad
	C = \frac{\brm{\vv{k}}^2-(\sqrt{s}-\omega)^2}{2A\brm{\vv{k}}}.
%	\label{BC}
}
For simplicity, the integration variable $\brm{\vv{k}}$ can be replaced by $\omega$, then:
\ba
	\frac{d\sigma_\gamma^{\rm hard}}{d\cos\theta}
	&=&
	\frac{\alpha^3}{4\pi s}
	\int\limits_\lambda^{\omega_{\rm max}}  d \omega \brm{\vv{k}}
	\int\limits_0^\pi d\theta_k \sin\theta_k
	\nn\\
	&&
	\int\limits_0^{2\pi} d\varphi_k
	\frac{\brm{\vv{k_2}}}{\varepsilon_1 g(\varepsilon_2)}
	\sum_i R_i  {\rm\Theta_P},
	\label{eq.HardPhysicKinematic}
\ea
where $\omega_{\rm max}$ is the maximum energy of the Bremsstrahlung photon.
$\lambda$ is conserved in the expression of the integrand, i.e., for example,
$\brm{\vv{k}} = \sqrt{\omega^2-\lambda^2}$.

The radiative invariants in $\sum_i R_i$ are expressed in terms of energies and angles as:
\eq{
	z_1 &= 2\omega E_1 + 2 \brm{\vv{k}} \brm{\vv{p_1}} \cos\theta_5,
	\nn\\
	v_1 & = 2 \omega E_2 - 2 \brm{\vv{k}} \brm{\vv{p_2}} \cos\theta_5,
	\nn\\
	z &= 2 \omega \varepsilon_2 + 2 \brm{\vv{k}} \brm{\vv{k_2}} A,
	\nn\\
	v &= 2 \omega (\sqrt{s}-\varepsilon_2) - 2\brm{\vv{k}} \brm{\vv{k_2}} A,
	\nn
}
where $E_{1,2}$ are the energies of the initial proton and antiproton in c.m.s.

%----------------------------------------------------
\section{Numerical analysis}
\label{sec.Numerical}
%----------------------------------------------------
%----------------------------------------------------

On the basis of the calculated cross section, two developed computer codes have been developed, namely a cross section calculator and a proper Monte Carlo event generator. In these programs, the user has full flexibility to define the kinematic region, set the values of the different parameters (fictitious photon mass, softness threshold, maximum energy of Bremsstrahlung photon, etc.), and also the possibility of including or excluding each of the radiative corrections terms, as long as the cancelation of divergencies is not violated. The latter feature will make both programs ideally suited for radiative corrections studies, including an accurate comparison of the results that will be obtained in PANDA  with other experiments.

In the following subsections we present the numerical results for radiative corrections using the formulas obtained in the previous sections.

%----------------------------------------------------
\subsection{Virtual and soft real photon emission}
\label{sec.numvirtsoft}
%----------------------------------------------------

\begin{table*}
\caption{\label{table.VirtSoft} Results for $s=5.08~\GeV^2$.
	Emission angle of electron (column I),
	fictitious photon mass (column II),
	Born cross section from (\ref{eq.BornWithFF}) (column III),
	virtual relative correction $\delta_V$ defined in (\ref{eq.delta}) and evaluated using
	(\ref{eq.BornVirt}) (column IV),
	soft real photon relative correction $\delta_\gamma$ defined in (\ref{eq.delta}) and evaluated using
	soft photon approximation by summing (\ref{eq.SoftISR}), (\ref{eq.SoftFSR}) and (\ref{eq.SoftOdd})
	(column V),
	sum of virtual corrections and soft photon emission (column VI). }
\begingroup
\renewcommand{\arraystretch}{1.3} % Default value: 1
\begin{tabular}{|clcccc|}
\hline
    $ \theta $ & $\lambda/\sqrt{s}$ & $\sigma_B$ & RC(virtual)  & RC(soft)        & RC(total)					\\
     (deg)     &                    &    ($\pb$) &  $\delta_V$  & $\delta_\gamma$ &  $\delta_V + \delta_\gamma$ \\
\hline   
$30$ & $10^{-6}$ & 22254.3  &       -0.52759    &    0.34791   &     -0.17968\\
         & $10^{-5}$ & 22254.3  &       -0.37606    &    0.19638   &     -0.17968\\
         & $10^{-4}$ & 22254.3  &       -0.22454    &    0.04486   &     -0.17968\\
         & $10^{-3}$ & 22254.3  &       -0.07301    &   -0.10667  &     -0.17968\\
\hline 
$60$  & $10^{-6}$ & 20478.6 &       -0.58627    &     0.38594    &     -0.20033\\
          & $10^{-5}$ & 20478.6 &       -0.42455    &     0.22422    &     -0.20033\\
          & $10^{-4}$ & 20478.6 &       -0.26283    &     0.06250    &     -0.20033\\
          & $10^{-3}$ & 20478.6 &       -0.10111    &   -0.09922    &     -0.20033\\
\hline                                      
$90$  & $10^{-6}$ & 19590.7 &        -0.65664   &    0.43191     &   -0.22474\\
          & $10^{-5}$ & 19590.7 &        -0.48275   &    0.25802     &   -0.22474\\
          & $10^{-4}$ & 19590.7 &        -0.30886   &    0.08412     &   -0.22474\\
          & $10^{-3}$ & 19590.7 &        -0.13497   &   -0.08977    &   -0.22474\\
\hline
$120$  & $10^{-6}$ & 20478.6     &  -0.72683   &       0.47788  &     -0.24895\\
            & $10^{-5}$ & 20478.6     &  -0.54077   &       0.29181  &     -0.24895\\
            & $10^{-4}$ & 20478.6     &  -0.35470   &       0.10575  &     -0.24895\\
            & $10^{-3}$ & 20478.6     &  -0.16864   &      -0.08031 &     -0.24895\\
\hline
$150$  & $10^{-6}$ & 22254.3          &  -0.78518   &      0.51591   &     -0.26928\\
            & $10^{-5}$ & 22254.3          &  -0.58893   &      0.31965   &     -0.26928\\
            & $10^{-4}$ & 22254.3          &  -0.39267   &      0.12339   &     -0.26928\\
            & $10^{-3}$ & 22254.3          &  -0.19641   &     -0.07287  &     -0.26928\\
\hline
\end{tabular}
\endgroup
\end{table*}

First we prove that the sum of virtual corrections with soft photon emission contribution
is free of infrared divergence. We regularize this infrared divergence by attributing to the photon 
a fictitious small mass $\lambda$. Numerically it is convenient to take this quantity small with respect to
the total energy of the process, i.e. say $10^{-6} < \lambda/\sqrt{s} < 10^{-3}$.
Table~\ref{table.VirtSoft} gives the results of first order RC calculation at $s=5.04~\GeV^2$, 
for different emission angles of electron, $\theta$, and with the threshold for soft photons $\Delta E/E = 1\%$.
The table shows the stability of the calculation on the fictitious photon mass $\lambda$. 
The calculation is very stable for any value of $\lambda$ on four orders of magnitude.
However, when $\lambda$ is too large (say, $\lambda > 10^{-4} \sqrt{s}$)
the soft photon approximation becomes invalid and may lead to unphysical results,
such as negative radiative cross sections.

%The total effect on the cross section is illustrated in
%Fig.~\ref{fig.TotalRcVsCosTh}, where the values of FFs, $\brm{G_E} = 0.16$ and $\brm{G_M} = 0.12$ have been
%taken from the vector meson dominance model \cite{Iachello:2004aq} at $s = 5.08~\GeV^2$ for the Born cross section.
%It has been verified that other FFs parametrizations do not change essentially the radiative corrections.
%The radiative cross section has been implemented in a Monte-Carlo program. The corresponding angular distribution
%of electron and positron generated events is illustrated in Fig.~\ref{fig.MC}.
%While a forward--backward asymmetry is created by the radiative events, when the charge of the final
%particle is selected, the symmetry of the total distribution is preserved.

\begin{figure}[h]
\includegraphics[width=0.52\textwidth]{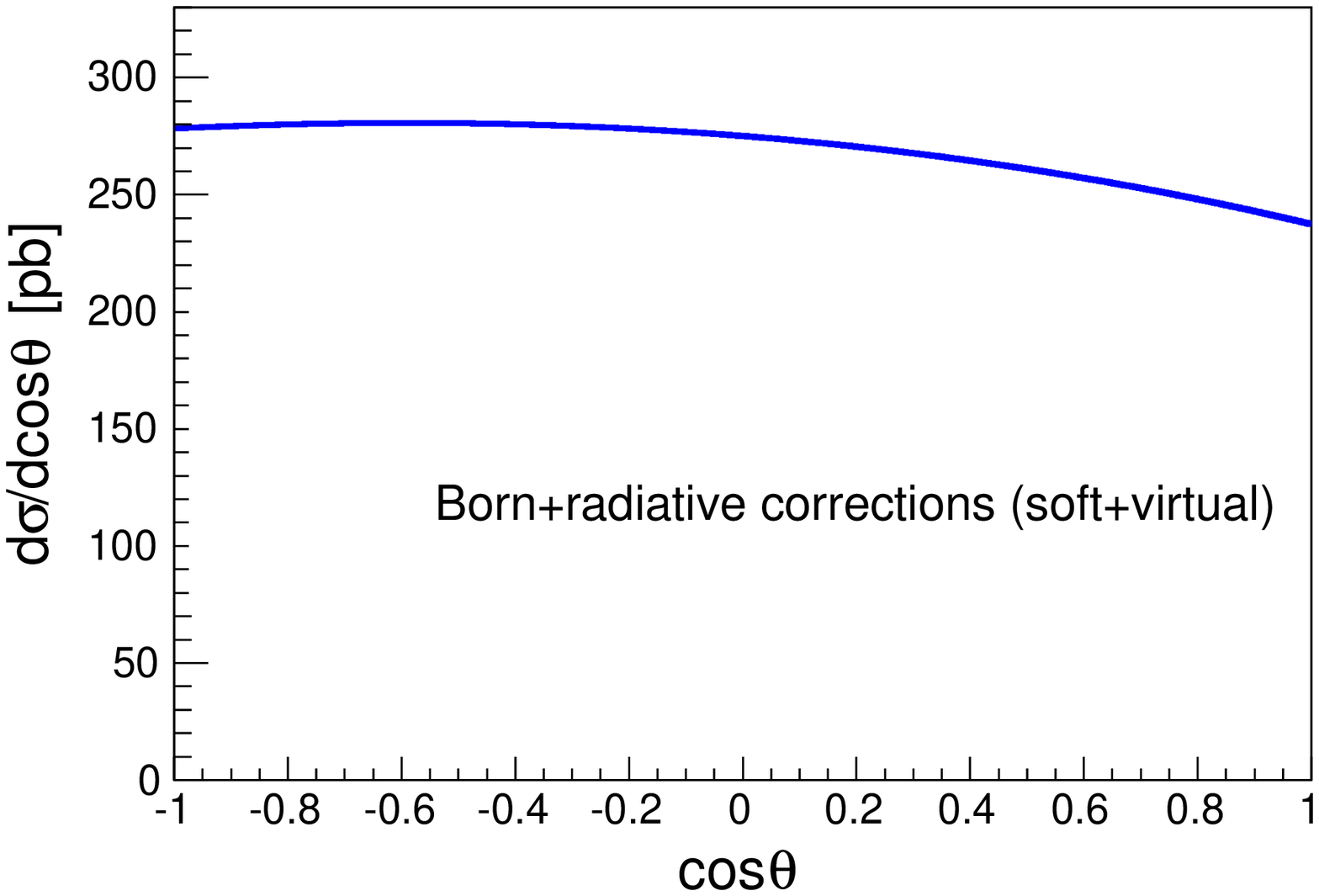}
\includegraphics[width=0.52\textwidth]{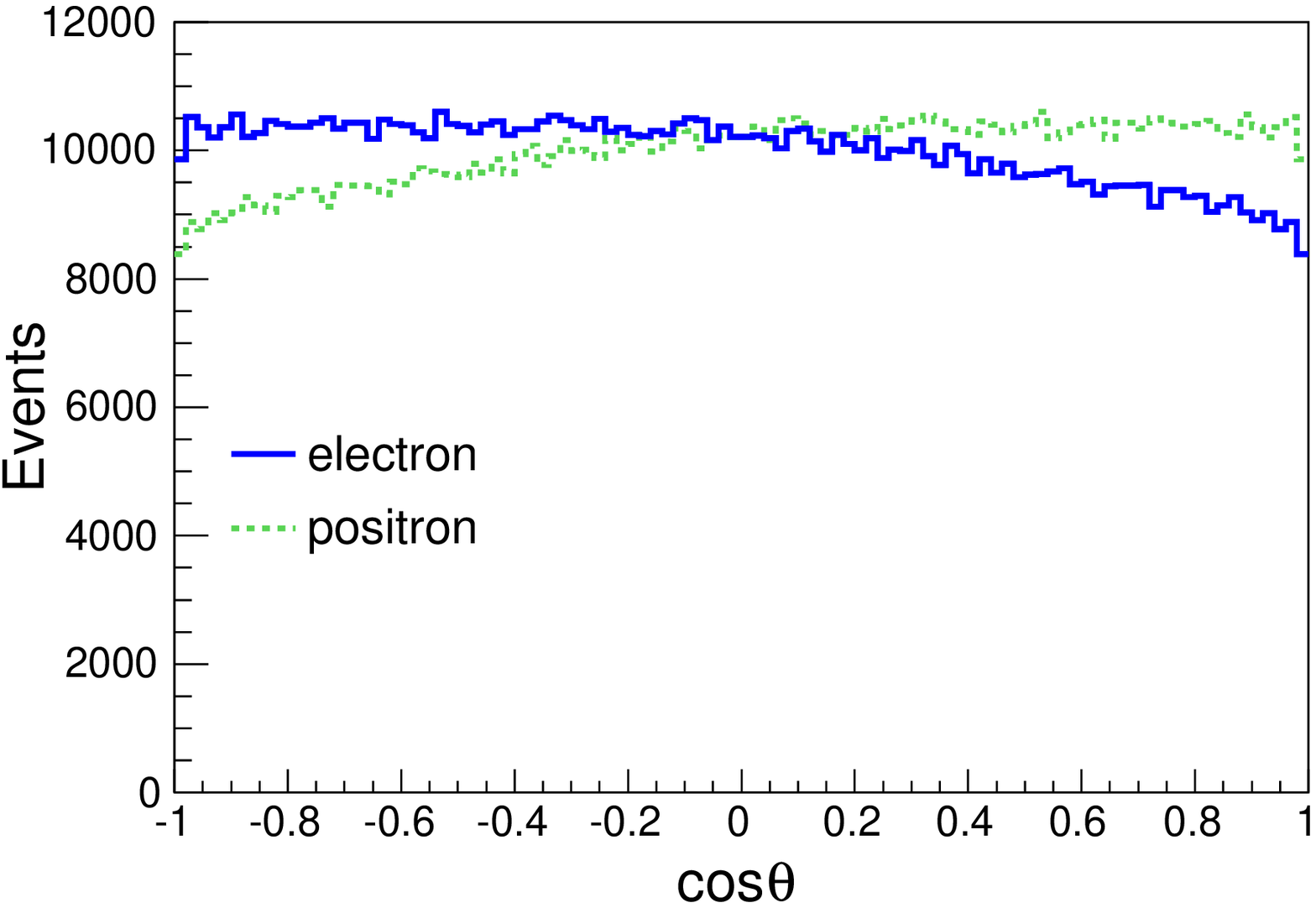}
\caption
{
Differential cross section as a function of $\cos\theta$ (top)  for a sample of $10^6$  generated events (top) for the process $\bar p p \to e^+ e^-$ including first order virtual and soft-real photon emissions,  for $s$=~5.08 GeV$^2$, with $|G_E|=0.16$ and $|G_M|=0.12$. The soft photon threshold is set to $\Delta E/E=1\%$. Bottom: corresponding number of events: the blue solid line (green dashed line)  describes the electron (positron) $\cos\theta$ distributions. 
}
\label{fig:VSevents}
\end{figure}

Figure~\ref{fig:VSevents} (left) displays the  radiative (virtual corrections and soft photon emission) differential cross section  of the process $\bar p p \to e^+ e^-$ as a function of $\cos\theta$ at  $s = 5.08$ GeV$^2$, corresponding to an antiproton momentum $p = 1.5$ GeV in the Lab frame. This value has been chosen as it will be the lowest available at PANDA. The values of the FFs, $|G_E|=0.16$ and $|G_M|=0.12$, are taken from the vector meson dominance model \cite{Iachello:2004aq}. The corresponding angular distributions of the electron and the positron for a sample of $10^6$ generated events are shown in Fig.~\ref{fig:VSevents} (right). 

\begin{figure}[h]
\includegraphics[width=0.52\textwidth]{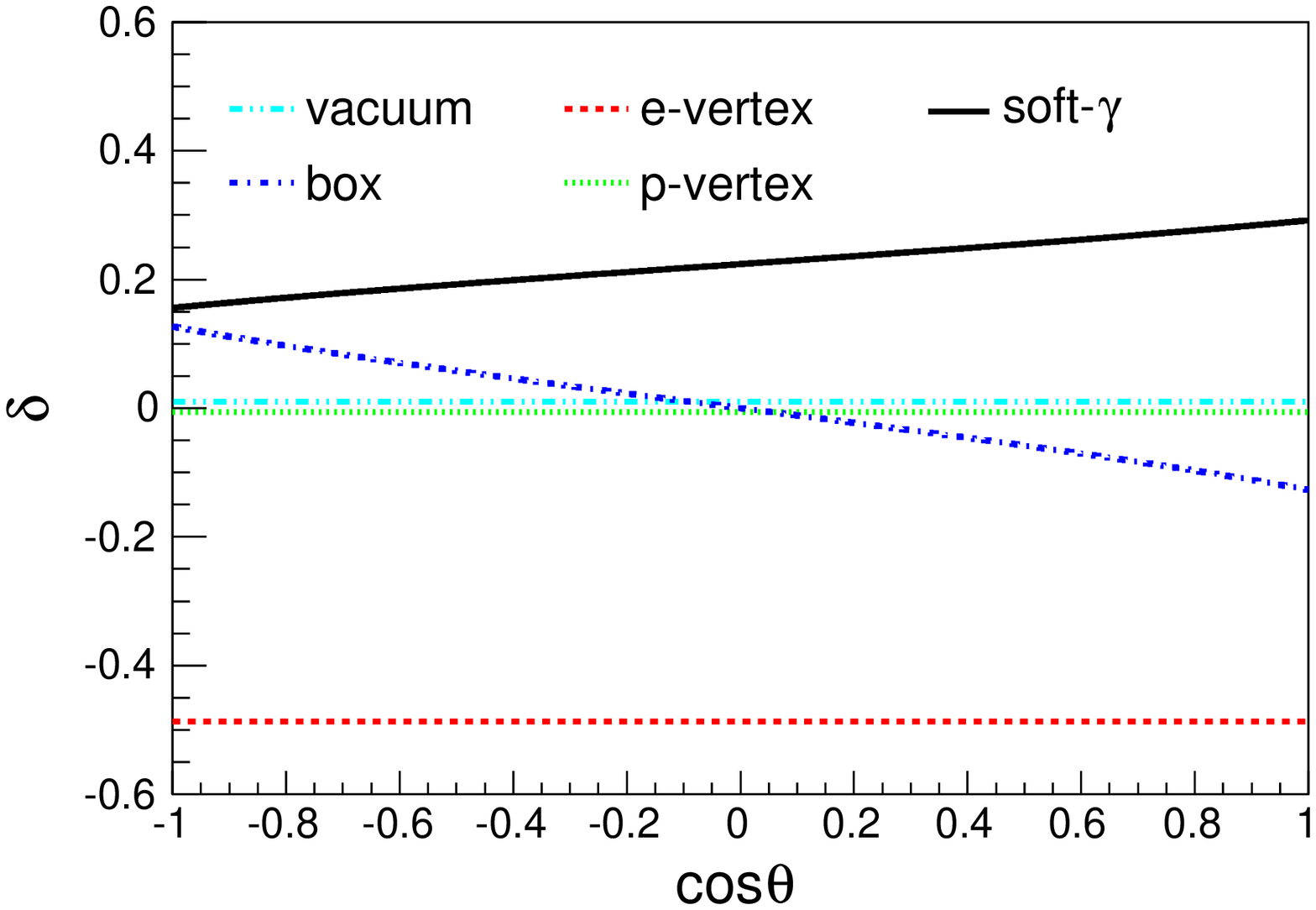}
\caption
[\protect{}]
{First order radiative corrections: soft photon emission (black solid line), virtual emission at electron vertex (red dashed line), virtual emission at the proton vertex (green dotted line), box diagrams (blue dashed-dotted line), vacuum polarization (Cyan dashed-dotted-dotted line), as a function of the electron c.m.s. production angle, for  $s$=~5.08 GeV$^2$ and $\Delta E/E=1\%.$
}
\label{fig.DeltaComponentsVsCosTh}
\end{figure}

In Fig.~\ref{fig.DeltaComponentsVsCosTh}, the different terms contributing to the virtual corrections and the contribution from the soft photon emission are plotted versus $\cos\theta$ for $s $=5.08~\GeV$^2$. The largest correction is the vertex correction from the electron vertex, that is constant in the angle. Then, the largest effects are given by the charge odd terms, i.e., the box diagram and the soft photon emission, that induce opposite forward--backward asymmetry. One can see that the vacuum polarization, the virtual corrections from the lepton and hadron vertices are $\theta$ independent, and therefore induce only a rescaling of the differential cross section.

%The forward--backward asymmetry in the $\cos\theta$ distributions is induced by the charge %odd terms, i.e., the box diagram and the soft photon emission. The vacuum polarization, the %virtual corrections from the lepton and hadron vertices are $\theta$ independent, and %therefore induce only a rescaling of the differential cross section.

\begin{figure}[h]
    \includegraphics[width=0.45\textwidth]{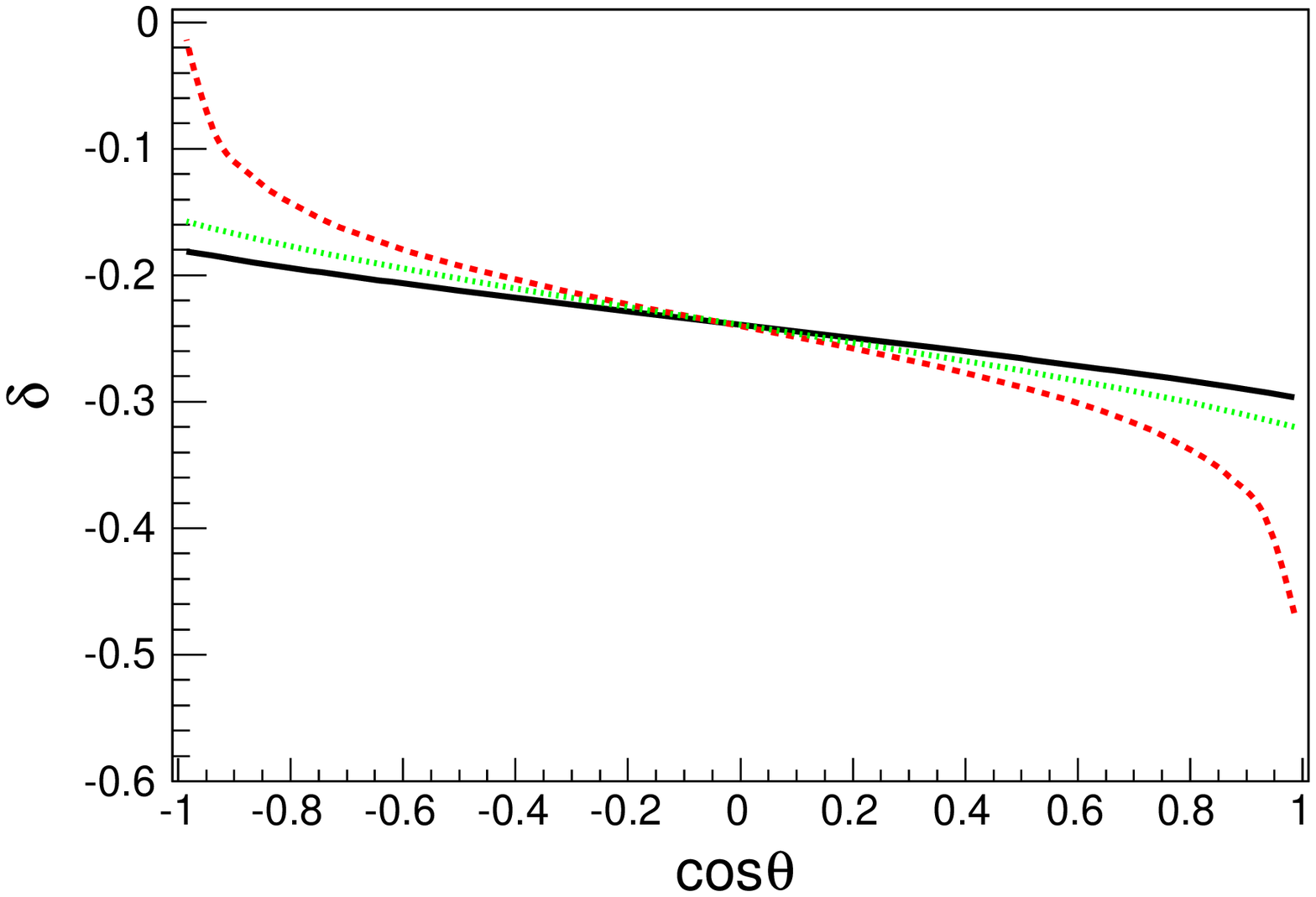}\\
\includegraphics[width=0.45\textwidth]{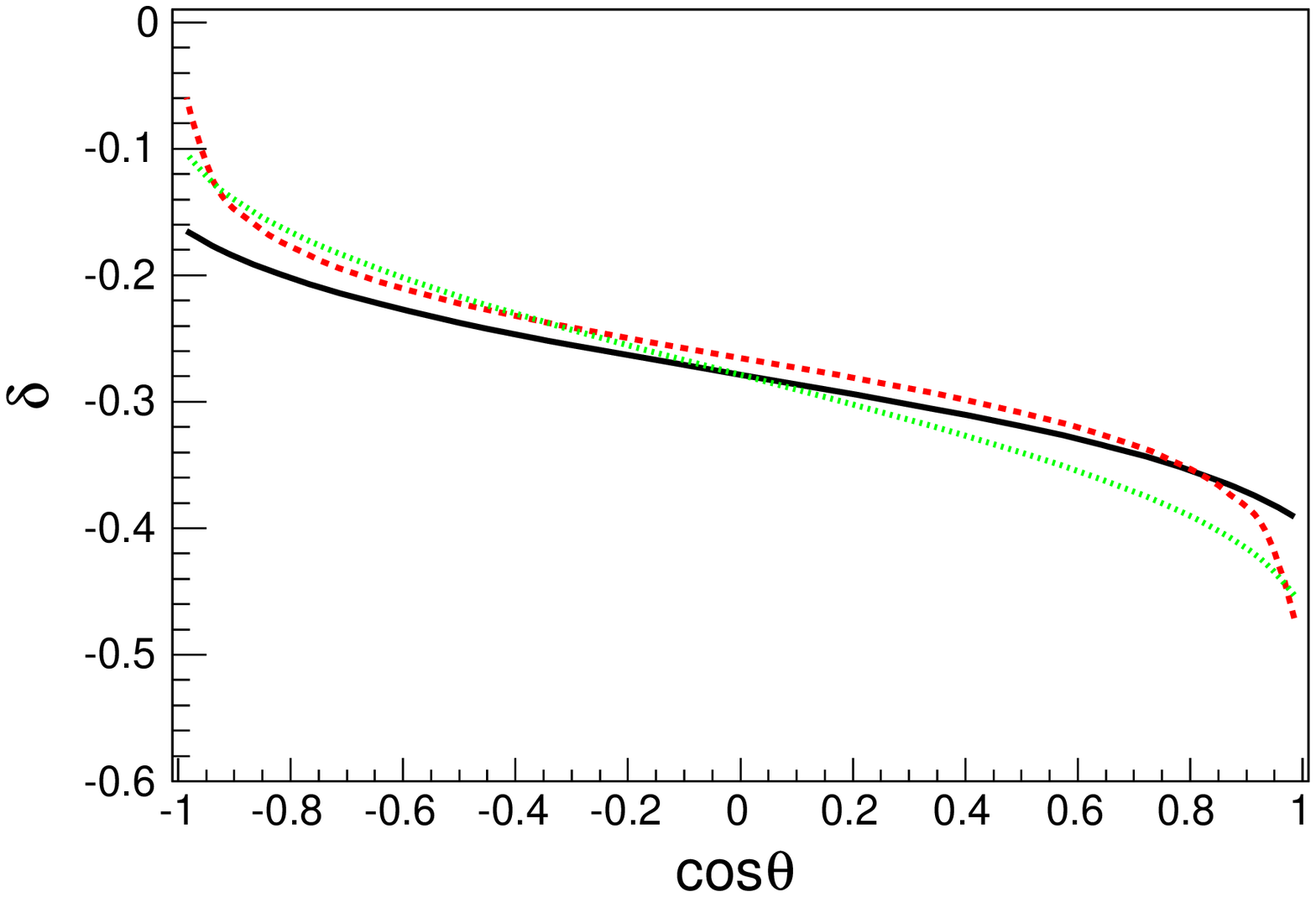}
\caption{ Soft real and virtual radiative corrections for $\omega/E=1\%$ at $s$=5.08\ GeV$^2$ (top) and  $s$=12.9 GeV$^2$ (bottom).  Calculations are - from \cite{Ahmadov:2010ak} (red dashed line)  - replacing the soft $ep$ interference from \cite{VandeWiele:2012nb} (green dotted line)
- and from \cite{Berends:1973fd} (black solid line). }
\label{Fig:Compare}
\end{figure}

The formulas without approximations from Ref. ~\cite{Ahmadov:2010ak}, 
have been implemented, except for Eq.~(27) of ~\cite{Ahmadov:2010ak}. The soft $ep$ interference has been 
calculated from Ref.~\cite{Berends:1973fd}. The difference between the exact 
first order calculation and the approximated one for the soft $ep$ interference are 
mostly due to the approximation $s\sim-t\sim-u\gg M^2$ in Eq.(27) of Ref.~\cite{Ahmadov:2010ak}, 
which does not hold  in the backward and forward angular regions. The results are reported in 
Fig.~\ref{Fig:Compare} for $\omega/E=1\%$ at $s$=5.08\ GeV$^2$ (top) and $s$=12.9 GeV$^2$ (bottom). 
These figures illustrate the effect of neglecting the proton mass and of 
accounting for the central angular region only.

%----------------------------------------------------
\subsection{Virtual, soft and hard real photon emission}
\label{sec.numreal}
%----------------------------------------------------

\begin{table*}
\caption{ \label{table.VirtSoftHard} The $\lambda$-independence of total relative corrections to cross section including
hard photon emission at different scattering angles $\theta$ and maximum Bremsstrahlung photon energies $\omega_{\rm max}$.
$V$ stands for virtual corrections (extracted from (\ref{eq.BornVirt})),
$R$ is for real photon emission contribution according to (\ref{eq.HardPhysicKinematic}).}
\begingroup
\renewcommand{\arraystretch}{1.3} % Default value: 1
\begin{tabular}{|c|c|c|c|c|c|c|c|} \hline
 \multicolumn{1}{|c|}{~~$\theta$~~}
&\multicolumn{1}{|c|}{~~$\lambda/\sqrt{s}$~~}
&\multicolumn{3}{c|}{ $\omega_{\rm max} = 0.1  \sqrt{s}/2$ } 
&\multicolumn{3}{c|}{ $\omega_{\rm max} = 0.3  \sqrt{s}/2$ } \\ 
 \cline{3-8}
 (deg) & & $V$ & $R$ &  $V+R$ & $V$ & $R$ & $V+R$ \\
\hline 
30 &   $10^{-8 }$ &    -0.83064  &   0.76575   &-0.06490      & -0.83064  &0.82476  &-0.00589 \\
      & $10^{-7}$  &    -0.67912  &   0.61423   &-0.06489      & -0.67912 & 0.67323  &-0.00588\\
      & $10^{-6}$  &    -0.52759  &   0.46274   &-0.06485      & -0.52759 & 0.52177  &-0.00582\\
      & $10^{-5}$  &    -0.37606  &   0.31166   &-0.06440      & -0.37606 & 0.37066  &-0.00540\\
      & $10^{-4}$  &    -0.22454  &   0.16885   &-0.05568      & -0.22454 & 0.22329  &-0.00125\\
      & $10^{-3 }$ &    -0.07301  &   0.06534   &-0.00767      & -0.07301 & 0.10193  & 0.02892\\
 \hline                                                                
90   &   $10^{-8}$  &    -1.00443  &   0.91202  &-0.09241       &-1.00443  & 0.98130  &-0.02313\\
       &  $10^{-7}$  &    -0.83053  &   0.73813  &-0.09240       & -0.83053 & 0.80742  &-0.02312\\
       & $10^{-6 }$ &    -0.65664  &   0.56429  &-0.09235       &-0.65664  & 0.63357  &-0.02307\\
       & $10^{-5 }$  &    -0.48275  &   0.39084  &-0.09192       &-0.48275  & 0.46010  &-0.02265\\
       & $10^{-4 }$  &    -0.30886  &   0.22565  &-0.08321       &-0.30886  & 0.29027  &-0.01859\\
       & $10^{-3 }$  &    -0.13497  &   0.09976  &-0.03521       &-0.13497  & 0.14634  &  0.01137\\
 \hline                                                                        
150 &   $10^{ -8 }$ &    -1.17770 & 1.05818    &-0.11952       &-1.17770 & 1.13727   &-0.04043\\
       & $10^{-7}$   &    -0.98144 & 0.86192    &-0.11952       &-0.98144 & 0.94102   &-0.04042\\
       & $10^{-6}$   &    -0.78518 & 0.66571    &-0.11948       &-0.78518 & 0.74480   &-0.04039\\
       & $10^{-5 }$  &    -0.58893 & 0.46987    &-0.11906       &-0.58893 & 0.54895   &-0.03998\\
       & $10^{-4 }$  &    -0.39267 & 0.28229    &-0.11038       &-0.39267 & 0.35665   &-0.03602\\
       & $10^{-3 }$  &    -0.19641 & 0.13402    &-0.06239       &-0.19641 & 0.19014   &-0.00627\\
\hline                
\end{tabular} 
\endgroup                                         
\end{table*}

Next we prove the $\lambda$-dependence cancellation when including hard photon emission.
In Table~\ref{table.VirtSoftHard} the virtual corrections and the sum of soft and hard photon emission (using Eq.  (\ref{eq.HardPhysicKinematic}) are shown. The stability of the sum with respect to $\lambda$ is worse than in Table~\ref{table.VirtSoft}, due to the Monte Carlo multidimensional integration of hard photon emission. Howevere, the stability of
the numerical results including hard photon remains at the $1\%$ level or better.

\begin{figure}
	\includegraphics[width=0.4\textwidth]{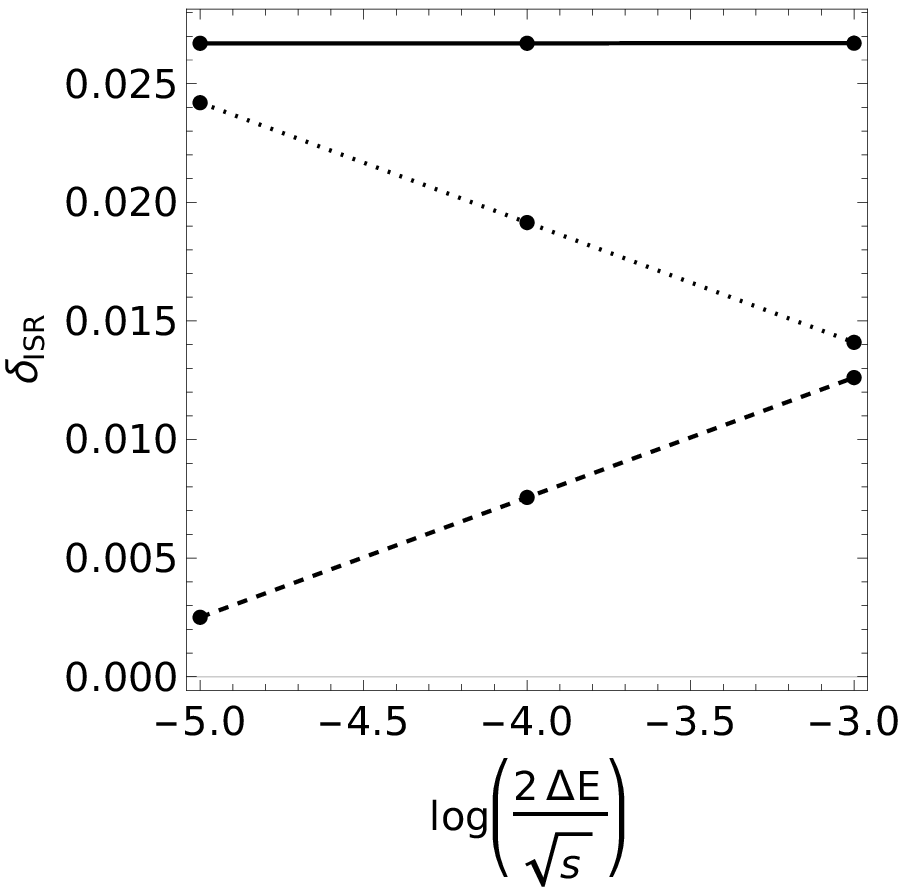}
	\includegraphics[width=0.4\textwidth]{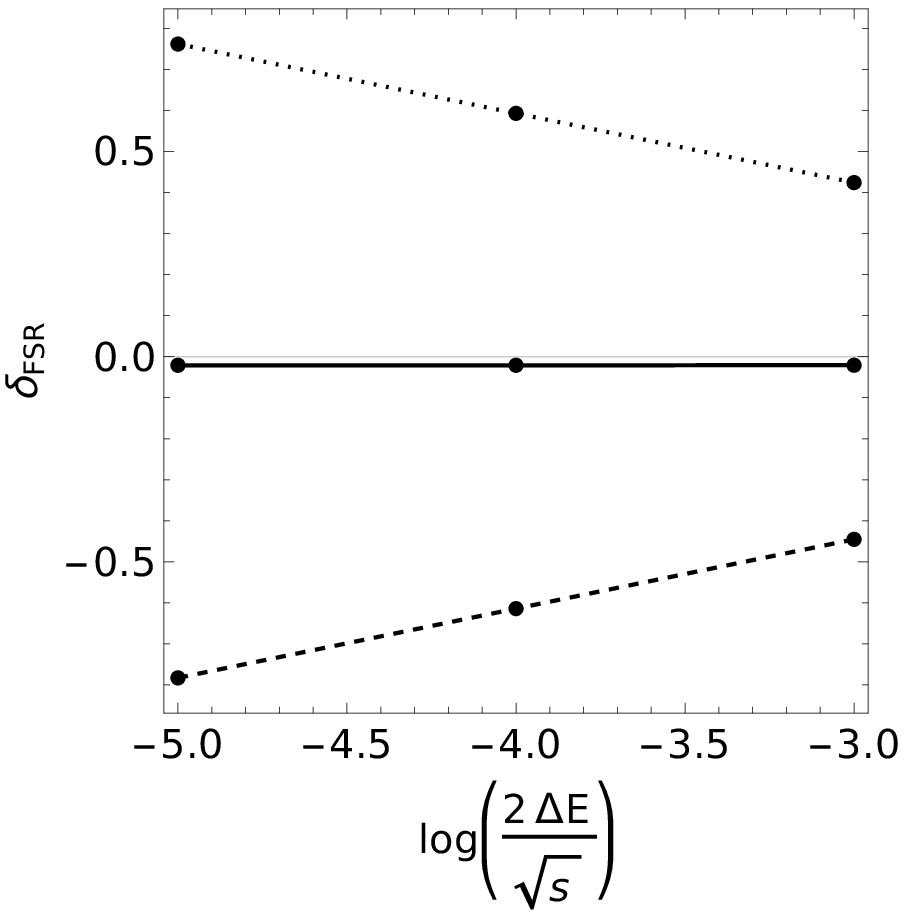}
	\includegraphics[width=0.4\textwidth]{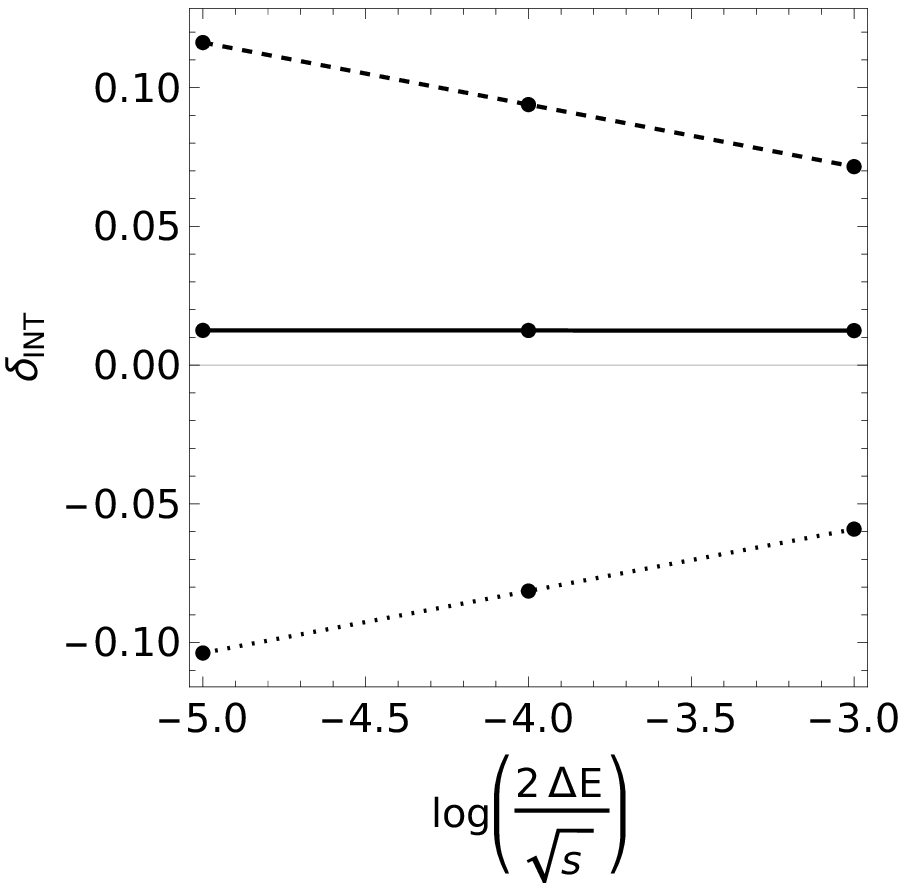}
\caption{\label{fig.OmegaCancellation} Illustration of the cancellation of the soft photon parameter $\Delta E$ in the sum of soft and hard photon emission at $s = 5.08~\GeV^2$. The dashed line is the sum of virtual corrections with emission of soft photons with energy below $\Delta E$. The dotted line is the hard photon emission with energy larger then $\Delta E$. The solid line is the sum of these two contributions. From top to bottom: ISR, FSR and INT corrections.}
\end{figure}

Finally we show that the total radiative corrections including virtual, soft and hard photon emission do not
depend on the soft photon parameter $\Delta E$ which delimits the regions of soft and hard photon emission.
In Fig.~\ref{fig.OmegaCancellation} the  three components charge-even : ISR, FSR and their charge-odd interference of the radiative corrections at $s$ = 5.08~\GeV$^2$ are illustrated, from top to bottom.
The dashed lines are the sum of virtual corrections with the corresponding soft photon emission with photon energy 
$\omega < \Delta E$. The boson self energy contributions are included in the upper plot.
The dotted lines correspond to  hard photon emission with photon energy $\omega > \Delta E$.
SThe solid lines are the sum of these two contributions. One can see that when scanning over a wide range for $\Delta E$,
the soft and hard photon emissions show a $\Delta E$-dependence while their sum does not.

\begin{figure*}[h!]
 \begin{center}
% \captionsetup[subfloat]{position=top,labelformat=empty}
  \subfloat[\label{ff1}]{\resizebox{0.5\textwidth}{!}{ \includegraphics{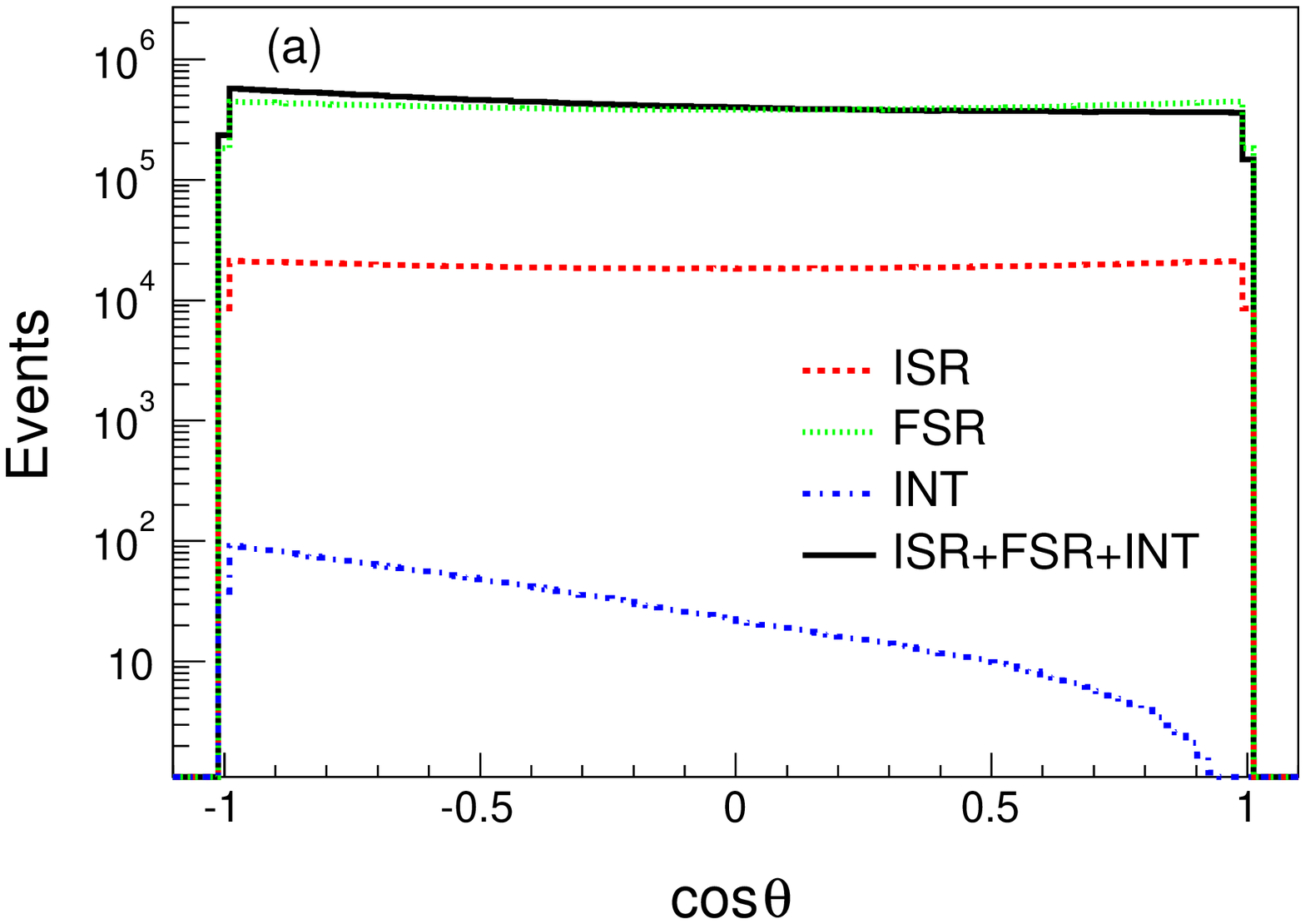} } }
  \subfloat[\label{ff2}]{\resizebox{0.5\textwidth}{!}{ \includegraphics{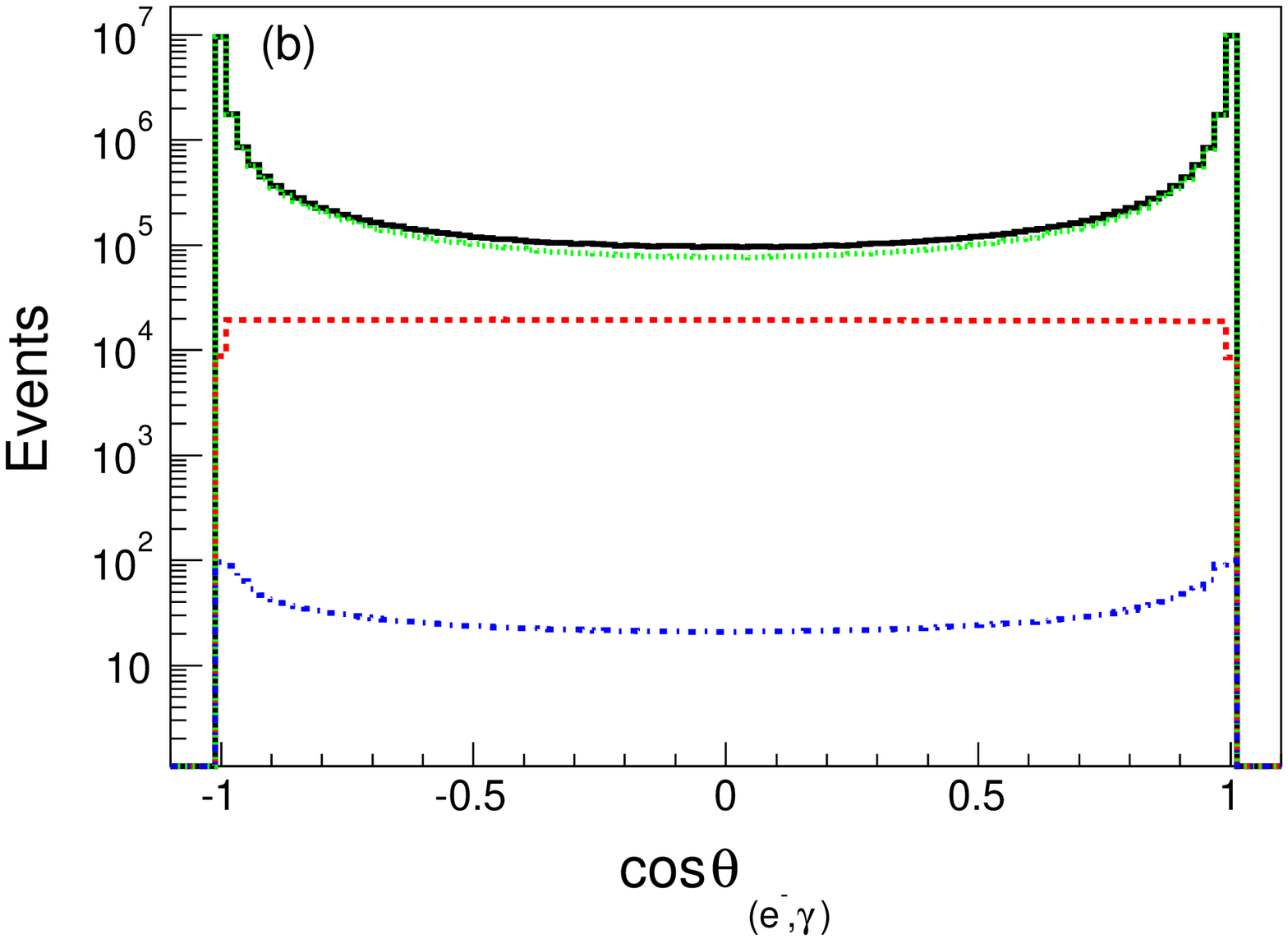} } }\\
 \subfloat[\label{ff3}]{\resizebox{0.5\textwidth}{!}{ \includegraphics{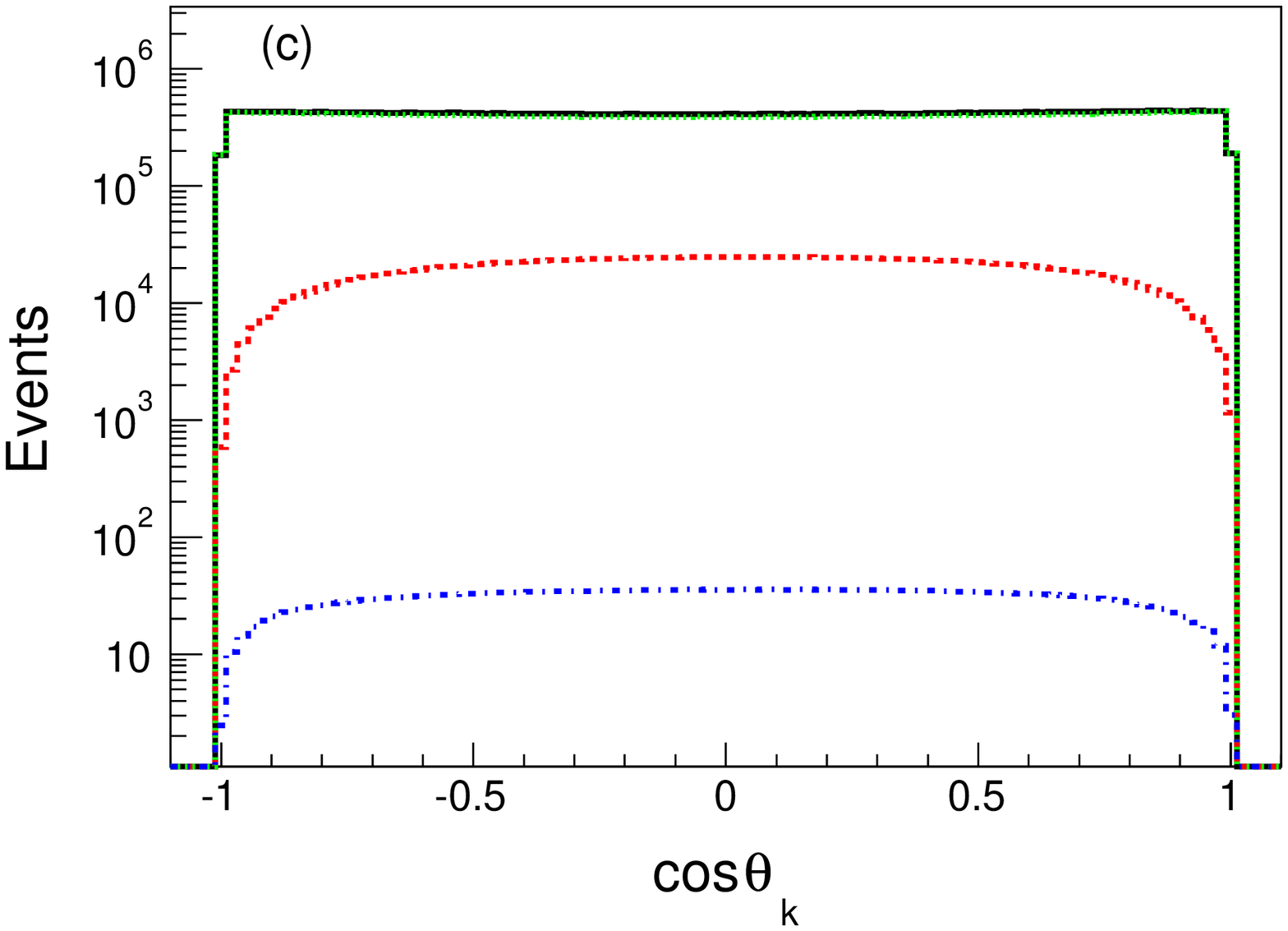} } }
  \subfloat[\label{ff4}]{\resizebox{0.5\textwidth}{!}{ \includegraphics{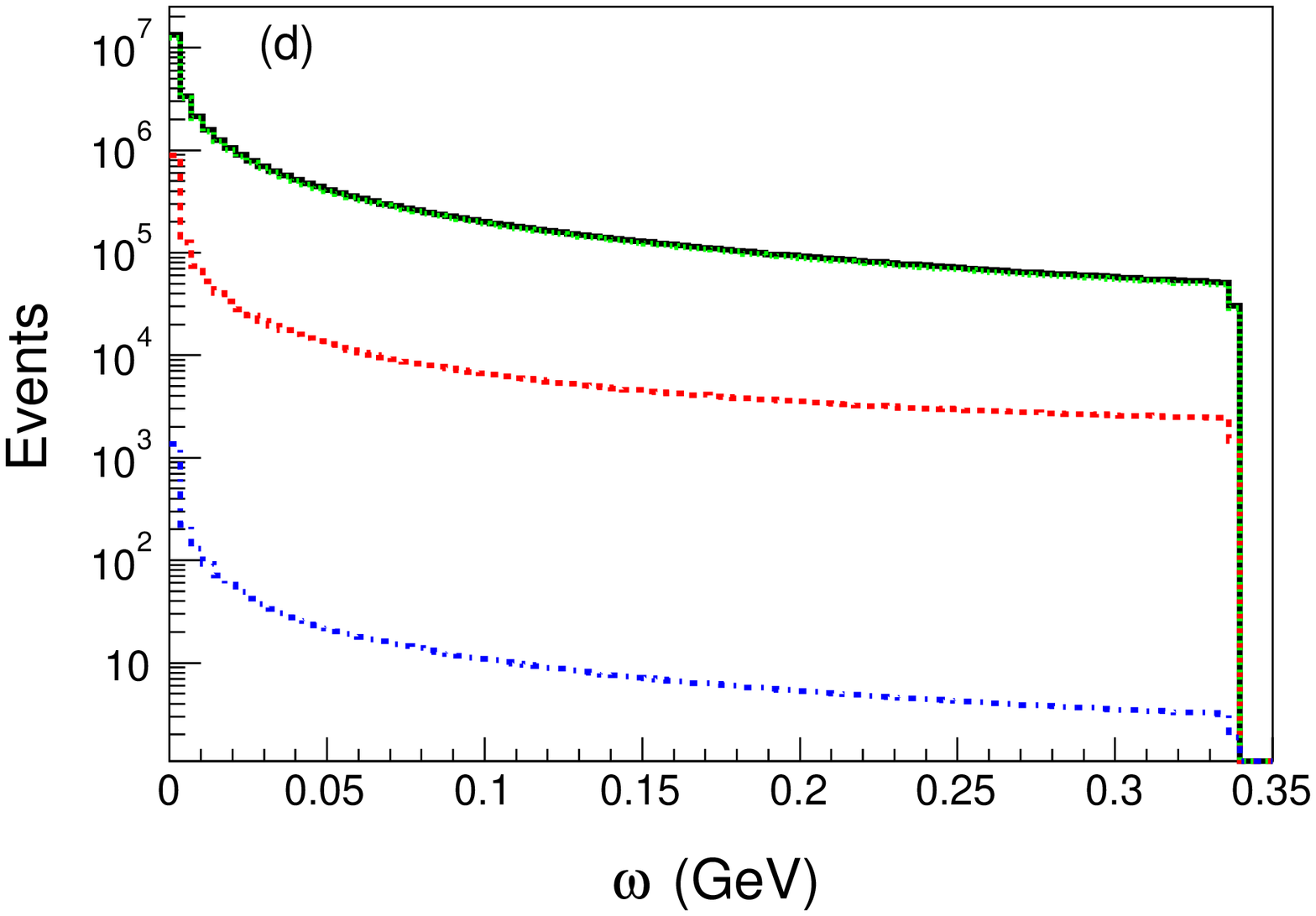} } }\\
 \subfloat[\label{ff5}]{\resizebox{0.5\textwidth}{!}{ \includegraphics{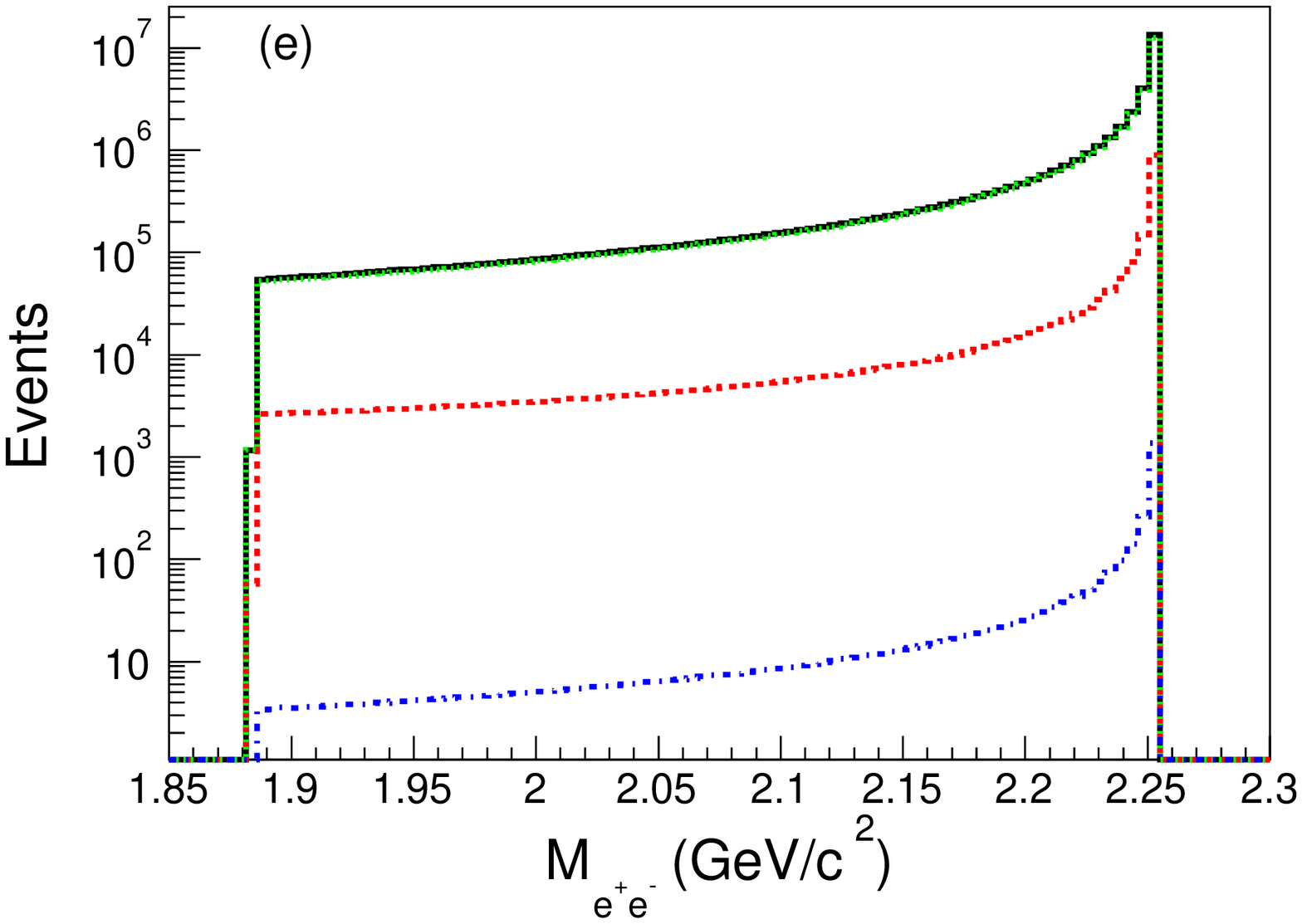} } }
  \subfloat[\label{ff6}]{\resizebox{0.5\textwidth}{!}{ \includegraphics{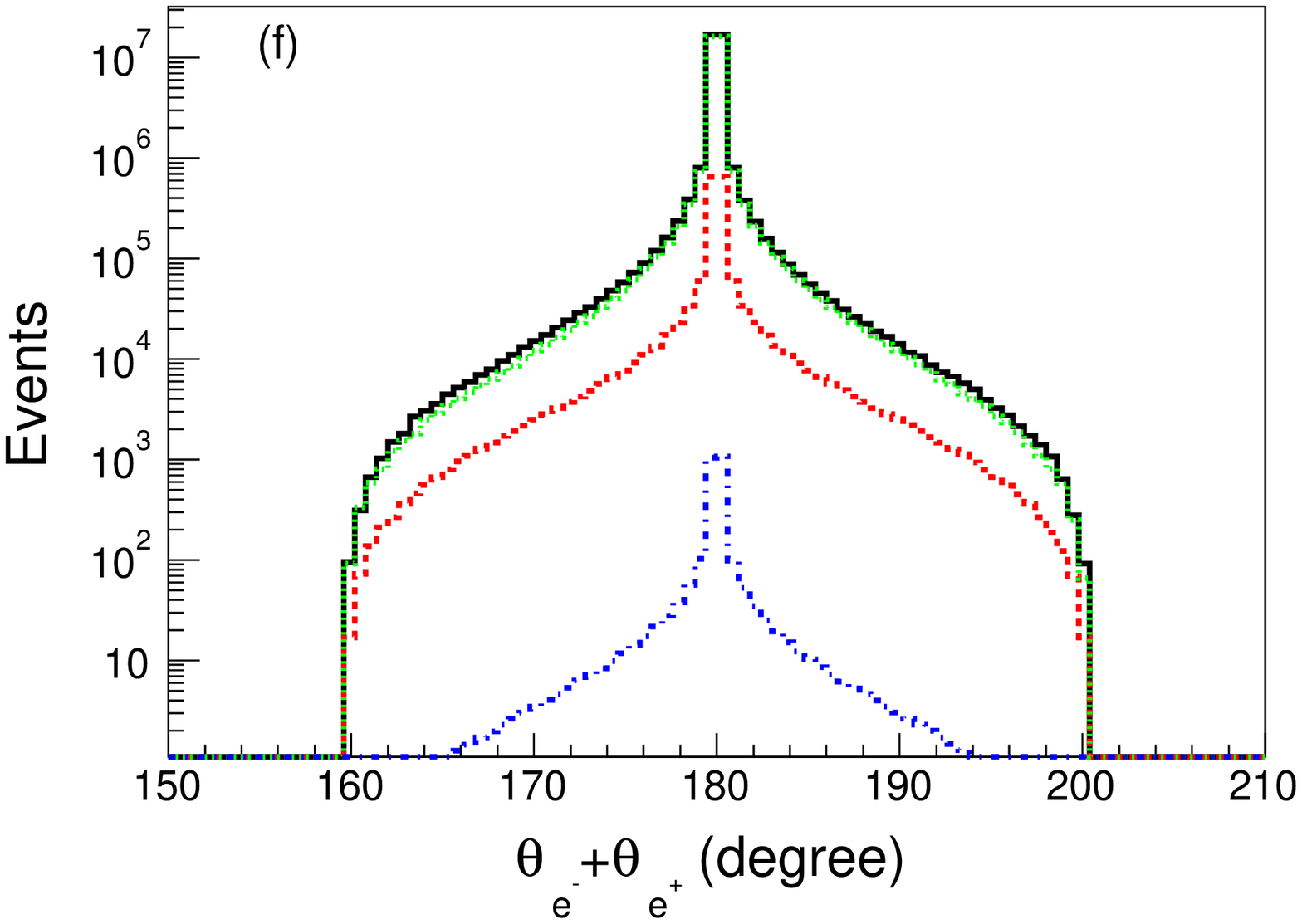} } }

\caption{\label{fig.HardEventGen}  Dependence of the five-fold Bremsstrahlung cross section (formula (\ref{eq.HardPhysicKinematic})) in the c.m.s, as output of the Monte Carlo event generator, on: (a)  the  electron angle $\cos\theta$, (b) the opening angle between the electron and the photon $\cos\theta_{(e^-,\gamma)}$,  (c)  the photon angle $\cos\theta_{k}$, (d) the photon energy $\omega$,  (e) the electron-positron invariant mass $M_{e^+e^-}$,  and (f) the sum of polar angles of the electron and positron. The red dotted histograms, the green dashed histograms, the dashed-dotted blue histograms and the black solid histograms describe the generated events from ISR, FSR, INT and ISR+FSR+INT, respectively. The c.m energy squared is $s=5.08$ GeV$^2$ ($p_{lab}=1.5$ GeV/c), the energy of the emitted photon is between $\lambda=10^{-5}\sqrt{s}/2$ and $w_{max}=0.3\sqrt{s}/2$. The number of generated events is $10^7$. The distributions are normalized to the numbers of the expected events according to their cross sections and a luminosity of 2 fb$^{-2}$. 
  }
\end{center}
\end{figure*}

Figure \ref{fig.HardEventGen} shows one-dimensional distributions of the Bremsstrahlung cross section (formula~(\ref{eq.HardPhysicKinematic})) in the c.m.s at $s=$5.08 GeV$^2$. The histograms are the outputs of the Monte Carlo event generator for the events corresponding to ISR (red dotted histograms) , FSR  (green dashed histograms) and INT (dashed-dotted blue histograms). The total contributions from  ISR+FSR+INT events are plotted as black solid lines.

The distributions of the electron polar angle $\cos\theta$ are illustrated in Fig.~\ref{fig.HardEventGen}a. One can see that the interference between photons emitted from initial and final states leads to a forward-backward asymmetry in $\cos\theta$ distribution.  The differential cross section of the reaction $\bar p \to e^+ e^- \gamma$ reaches its highest values at small photon polar angles relative to the direction of the final state electron and positron (Fig.~\ref{fig.HardEventGen}b). The distributions of the photon polar angle $\cos\theta_{k}$ are illustrated in Fig.~\ref{fig.HardEventGen}c. In the considered reference system, this angle is the opening angle between the photon and this initial state antiproton. Fig.~\ref{fig.HardEventGen}d shows the energy distributions of the emitted photon. Soft and hard  photon contributions are included. 

In these simulations, the photon energy range is taken from $\lambda=10^{-5}\sqrt{s}$ up to  $w_{max}=0.3\sqrt{s}/2$. The value of $w_{max}$ is typically above the energy resolution of the PANDA detector and can be used as an experimental cut to select the signal events $p \bar p \to e^+ e^- (\gamma)$ (e.g $\omega<w_{max}$. To suppress the background channels, one can also apply a cut on the invariant mass  $M_{e^+e^-}$ of the $e^+e^-$  system (Fig.~\ref{fig.HardEventGen}e), e.g. $a<M_{e^+e^-}<b$, where the values of $a$ and $b$  are chosen  in order  to achieve a high background suppression factor keeping a good signal efficiency \cite{Singh:2016dtf}. 

The $M_{e^+e^-}$ for the Born events ($p \bar p \to e^+ e^-$) corresponds to a c.m. s. energy $\sqrt{s}=2.25$ GeV. The radiative tail is due to hard photon emission. Another relevant kinematic variable  for the signal selection is the sum of the electron and positron polar angles  ($\theta_{e^+} + \theta $). The electron and positron are emitted back to back in the c.m.s  (($\theta_{e^+} + \theta $)=180$^{\circ}$). The tails  shown around $180^{\circ}$ (Fig.~\ref{fig.HardEventGen}f) are due to hard photon emission.

The differential cross section for the process $\bar p \to e^+ e^- \gamma$ presents some peaks that correspond to the situation when the emitted photon is collinear to the direction of the electron or positron. This leads to a reduction in the efficiency and in the accuracy of the Monte Carlo algorithm. To absorb these peaks and allow a fast event generation, the Importance Sampling method  \cite{Hammersley:19t4} has been used.  Jacobian transformations  of the two variables, the photon polar angle and energy, have been performed as described in Ref.~\cite{Caffo:1996mi}.

%----------------------------------------------------
\section{Conclusion}
\label{sec.Conclusion}
%----------------------------------------------------

The precise measurements of the TL electromagnetic FFs of the proton expected at the future PANDA experiment via the reaction $\bar{p}p \rightarrow e^+e^-$, require to take into account radiative corrections. Although several packages dealing with radiative corrections are available on the market, none of them is entirely suitable for PANDA. In this work,  the next-to-leading-order (NLO) $\bar{p}p \rightarrow e^+e^-$ differential cross section has been calculated in the point-like approximation, including both virtual and real corrections. 

Relying on the method developed in Ref. \cite{Ahmadov:2010ak}, the full set of calculated virtual corrections include vacuum polarisztion (with all leptons and a point-like pion in the photon loop), corrections to both electron and proton vertex, and two photon exchange. On the other hand, real corrections including initial and final state radiation, and the interference between them, have been re-calculated in this work, in the spirit of  the classic references \cite{Berends:1973tz} and \cite{Berends:1973fd} for the reactions $e^+e^- \rightarrow \mu^+\mu^-$. In the soft photon regime, infrared divergences from singular virtual diagrams are cancelled out with the corresponding real diagrams. On the other hand, the regularisation of infrared divergencies of the Bremsstrahlung cross section is based on the introduaction of a small photon mass as a parameter, which makes the calculation applicable to both the soft and hard photon regimes. The calculated cross section is at the basis of two computer codes which can be included in the PANDA physics analysis framework.

\section{Acknowledgments}
The authors thank  H.~Czyz for interest in this work and useful suggestions. Three of us (Yu.M. B., V.A. Z., and E. T.-G.) thank the Helmholtz Institute Mainz for warm hospitality and support. This work was supported by the German Ministry for 
Education and Research (BMBF) under grant number 
05P12UMFP9 and under grant number 05P19UMFP1.

%----------------------------------------------------

\appendix

%----------------------------------------------------
\section{Kinematical quantities in c.m.s expressed in terms of invariants}
\label{sec.KinematicsViaInvariants}
%----------------------------------------------------

The expression of the physical kinematical variables are given in terms of
the radiative invariants (see Table~\ref{table.Invariants}).
All formulas in this section are given neglecting the quantities $m$ and $\lambda$.

\begin{enumerate}

	\item
		The photon energy:
		\eq{
			\omega = \frac{z + v}{2\sqrt{s}}.
		}

	\item
		The photon momentum $\vv{k}$ polar angle:
		\eq{
			\cos\theta_k = \frac{z_1 - v_1}{2 \omega \sqrt{s - 4M^2}}.
		}

	\item
		The photon momentum $\vv{k}$ azimuthal angle:
		\eq{
			\cos\varphi_k = - \frac{A + \cos\theta \cos\theta_k}{\sin\theta \sin\theta_k},
		}
		where
		\eq{
			A = \frac{v - z - 2 \omega \br{2\varepsilon_2 - \sqrt{s}}}{4\omega \sqrt{\varepsilon_2^2 - m^2}}.
		}

	\item
		The electron energy:
		\eq{
			\varepsilon_2 =
			\begin{cases}
				\br{1-\frac{z}{s}} \frac{\sqrt{s}}{2},\qquad\qquad \mbox{if}\ v > z  \frac{s-z}{s+z},
				\\
				\br{1-\frac{v}{s-z-v}} \frac{\sqrt{s}}{2},\qquad \!\!\mbox{if}\ v < z  \frac{s-z}{s+z}.
			\end{cases}
		}

	\item
		The positron energy:
		\eq{
			\varepsilon_1 = \sqrt{s} - \varepsilon_2 - \omega.
		}

\end{enumerate}

\bibliography{Bibliorad2}

\end{document}